\documentclass{article}
\usepackage[utf8]{inputenc}
\usepackage{longtable}
\usepackage{booktabs}
\usepackage{graphicx}
\usepackage{appendix}
\usepackage{siunitx}
\usepackage{amssymb}
\usepackage{amsmath} 
\usepackage{comment}
\usepackage{hyperref}
\usepackage{listings}
\AtBeginDocument{}
\usepackage[capitalise,noabbrev]{cleveref}
\usepackage[utf8]{inputenc}
\usepackage[english]{babel}
\usepackage[nottoc,numbib]{tocbibind}
\usepackage{todonotes}
\usepackage{csquotes}
\usepackage{tablefootnote}
\usepackage{amssymb} 
\usepackage{pifont}
\usepackage{framed}

\usepackage{biblatex}[backend=biber, maxnames=30, style = unsrt]

\begin{document}

\newcommand{\btoprogram}{{\sc B2Program}}
\newcommand{\prob}{{\sc ProB}}
\newcommand{\simb}{{\sc SimB}}

\newcommand{\cmark}{\ding{51}}%
\newcommand{\xmark}{\ding{55}}%
\newcommand{\questionmark}{\textbf{?}}%

\begin{titlepage}
   \centering

    \hfil\includegraphics[width=8cm]{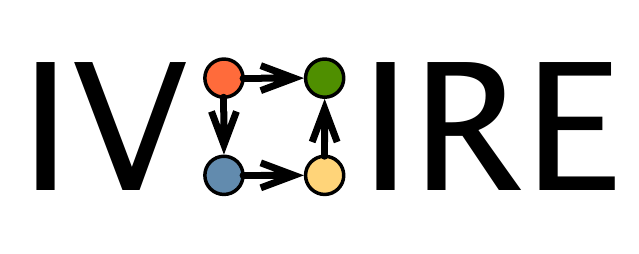}\newline
    \hfil\hfil\includegraphics[width=5cm]{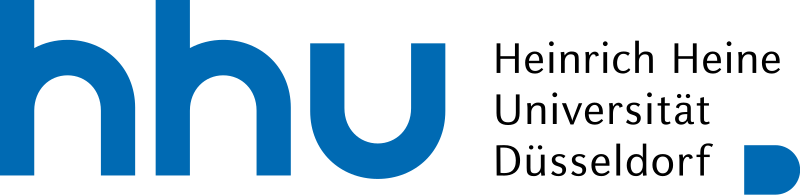}
    \hfil\hfil\hfil\hfil\includegraphics[width=4cm]{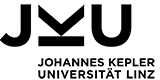}\newline
    \par\vspace*{4\baselineskip}
    {\huge \bfseries IVOIRE -  Deliverable D 1.1}\\[1.5cm]
\textsc{\Large{\today}\\[1.5cm]
 Version 1.2.0}\\[6.0cm]

Sebastian Stock, Fabian Vu, Atif Mashkoor, Michael Leuschel, Alexander Egyed

\end{titlepage}

\tableofcontents
\newpage

\section{Introduction}
\label{sec:introduction}
This document presents the report for D1.1 ("Classification of existing Validation Obligations and Tools") of the IVOIRE project.\\
Within the context of software engineering, it is vital to check whether a model meets its specification (verification) and requirements (validation).
Various formal techniques exist to verify a requirements model, such as theorem proving or model checking. However, to validate a requirements model, only a few formal techniques are at our disposal.
Verification checks that a system or piece of software meets its specification.
This means that the program is proven to be correct concerning its specification \cite{ideStdVerif}. {\bf Verification} answers the question: ``Are we building the software correctly?''
In contrast, {\bf validation} checks whether a model met the stakeholder's requirements \cite{ideStdVal}. It answers the question: ``Are we building the right software?''.

In general, verification ensures the absence of bugs in the software, e.g., absence of infinite loops, absence of integer overflows. 
Proof obligations (POs) have been introduced, e.g., for B \cite{abrial2005b} and Event-B  \cite{abrial2007rodin} to structure the verification process.
A successfully discharged PO ensures the absence of a safety hazard in a model.
Validation ensures the presence of certain features. 
Therefore, a successfully validated system contains desired behaviors, e.g., the ability to perform actions in a specific order.

Verification and validation may overlap, e.g., establishing the presence of a safety property can -- depending on the context -- be viewed either as verification or validation.
Let us consider a safety requirement ``The lift can only move when the door is closed.'' One may argue that checking this requirement is verification rather than validation.
However, the main goal is to check that the stakeholders' requirements are fulfilled. So, in this context, model checkers and theorem provers, which are primarily verification tools, can also be used for validation. 

Compared to verification, validation has received less attention historically in the formal methods community.
This is perhaps due to its unprovable nature, as suggested by Rushby:
``By their
very nature, the problems of validating top-level specifications or statements of assumptions do not lend themselves to definitive proof'' \cite{Rushby93:FAA}.
Nonetheless, as outlined by Jacquot and Mashkoor~\cite{role_validation} validation within a refinement-based development process is still a challenging task.

In this context, validation obligations (VOs)~\cite{mashkoor2021validation} were introduced to check the compliance of a model with its requirements in a refinement-based software development process.
While systematic validation of formal models is not a new concept in the formal methods community, the VO approach is a generic approach independent of a particular formal method or tool.

This report defines the term \emph{validation obligation}, and presents the formalization of VOs and the underlying techniques.
More concretely:
(1) it formalizes VOs, allowing a combination of tasks and techniques to validate requirements,
(2) it proposes semantics that clarifies the dependencies between the associated validation tasks, and 
(3) it provides a formal basis to trace validation tasks back to the requirement in a model.

VOs must therefore be resilient when more details are added to a model.
As requirements are typically expressed in natural language, they can be ambiguous or imprecise.
Although a VO aims to be a formal representation, some validation tasks may retain a manual and/or informal component.
For example, a certain VO may produce a visualization that still has to be inspected by a domain expert.

First, we will present the classification of requirements (see \cref{sec:requirements}).
In \cref{sec:vandv}, we will discuss the terminologies \emph{verification} and \emph{validation}.
Then, we define the term \emph{validation obligation} along with how associated \emph{validation tasks} are formalized and classified (see \cref{wording}). Here, we also present an overview of existing validation tasks for the modeling languages Alloy, ASM, B, Event-B, VDM, TLA+, Z, CSP, and Circus.
Furthermore, we will describe how VOs are used in a refinement-based software development process to validate requirements.
Finally, we will demonstrate the VO approach on a Traffic Light model in \cref{sec:demonstration}.

A glossary containing the basic terms can be found in  \cref{sec:glossary}.
This report also includes a list of publications in the context of the IVOIRE project \cref{sec:publications}. 

\section{Classification of Requirements}
\label{sec:requirements}




By definition according to the IEEE standard 729 \cite{ieee729}, a requirement is defined as follows:

\begin{enumerate}
\item A condition or capability needed by a user to solve a problem or achieve an objective.
\item A condition or capability that must be met or possessed by a system or system component to satisfy a contract, standard, specification, or other formally imposed document.
\item A documented representation of a condition or capability as in 1 or 2.
\end{enumerate}

This section describes how requirements are classified.
In general, they are separated into functional requirements, non-functional requirements, and domain requirements.
Furthermore, requirements could also be distinguished between user requirements, and system requirements. \cite{Sommerville10}

\subsection{Functional, Non-Functional, and Domain Requirements}

Functional requirements describe how the system should behave.
Regarding the general definition of a requirement, functional requirements match the definition of the first aspect. \cite{Sommerville10}

Thus, functional requirements include descriptions of safety properties, liveness properties, scenarios, probabilistic behaviors, timing behaviors.

In contrast, non-functional requirements describe measurements or constraints for the quality of the software system such as performance, reliability, maintainability, testability, scalability, or security.
Thus, non-functional requirements define criteria to evaluate those quality constraints or measurements. \cite{Sommerville10}

Taking a look at the definition of a requirement, non-functional requirements correspond to the second aspect.

Domain requirements are requirements that have been formulated from a domain expert's perspective.
Therefore, domain requirements can either be functional or non-functional. \cite{Sommerville10}

Concerning the domain-specific aspect, it might be necessary to refine or abstract the model, projecting on the domain expert's perspective.
Since the model is projected on a specific perspective, state space projection and refinement might play an important role during the validation.

\subsection{User Requirements and System Requirements}

Requirements could also be distinguished between user and system requirements.
User requirements are written from a stakeholder's perspective, describing the expectation of how a system should behave.
Therefore, user requirements usually describe how the user can interact with the model, and check the software's behavior.
In contrast, system requirements describe how software components interact with each other.
Thus, system requirements are rather architectural or structural. \cite{Sommerville10}

\section{Verification and Validation}
\label{sec:vandv}

Sometimes the terms verification and validation are used interchangeably. In our report, we use the following definitions.
Validation checks whether a model meets the stakeholders' requirements.
So, the main questions are: ``Are we building the right software?'' and ``Are the desired features present?''
In contrast, verification checks whether a model meets its specification. So, it tackles the questions: ``Are we building the software correctly?'', ``Are all safety constraints be enforced?''.

For example, verification ensures that there are no bugs in a model.
Properties that correspond to verification (and not validation) are, e.g., the absence of integer overflows or infinite loops.
Those properties could be verified by model checking or proving.
An example for validation is checking a
 a requirement described by a scenario.
This requirement can be validated by running the software with specific input to observe and check the behavior.
While verification and validation are different tasks, they also complement each other.
For example, well-definedness checking is classified as verification. However, while finding such an error by model checking, one could store the trace as a test for the corrected model (i.e., desiring the absence of well-definedness errors), which falls within the jurisdiction of validation.

\begin{figure}[t]
   \centering
   \includegraphics[scale=0.38]{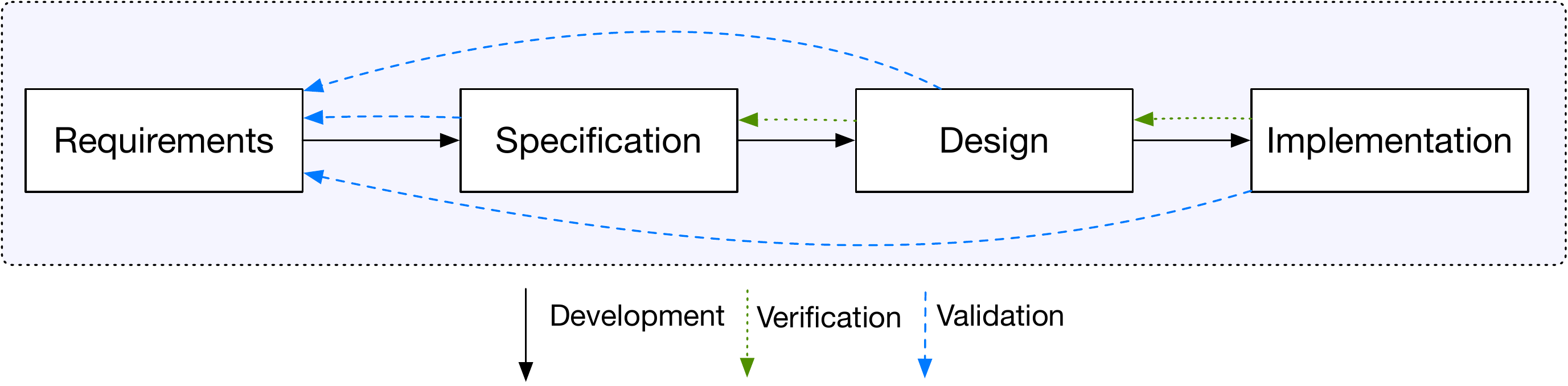}
   \caption{Role of Verification and Validation}
   \label{fig:vav}
\end{figure}

Consider~\cref {fig:vav} illustrating the role of verification and validation in software development.
Verification ensures consistency between specification, design, and implementation.
In contrast, validation ensures that the specification, design, and implementation fulfill the stakeholders' requirements.

One might argue that verification should precede validation as the reverse would waste time and effort: Why should someone try to validate the software's behavior although the software is incorrect? Nevertheless, this question could also be asked the other way around: Why should someone try to verify the software with the potential to notice that the software does not behave as desired in the end? While there is no definitive answer as it is a subjective question, our position is that since both verification and validation are equally important activities, therefore, should be given equal preeminence.

\section{Validation Obligations Approach}
\label{wording}

This section presents a formalization of validation obligations (VO) and a classification of validation tasks (VT).
As explained by Mashkoor et al.~\cite{mashkoor2021validation}, refinement plays an important role in the VO approach.
Therefore, the formal representation of VOs should also provide a basis to transform, refine or abstract VOs.
Furthermore, this work discusses how VTs are created and how VOs are integrated into the software development process.

\begin{center}
\begin{figure}[ht]
    \centering
    \includegraphics[width=12cm]{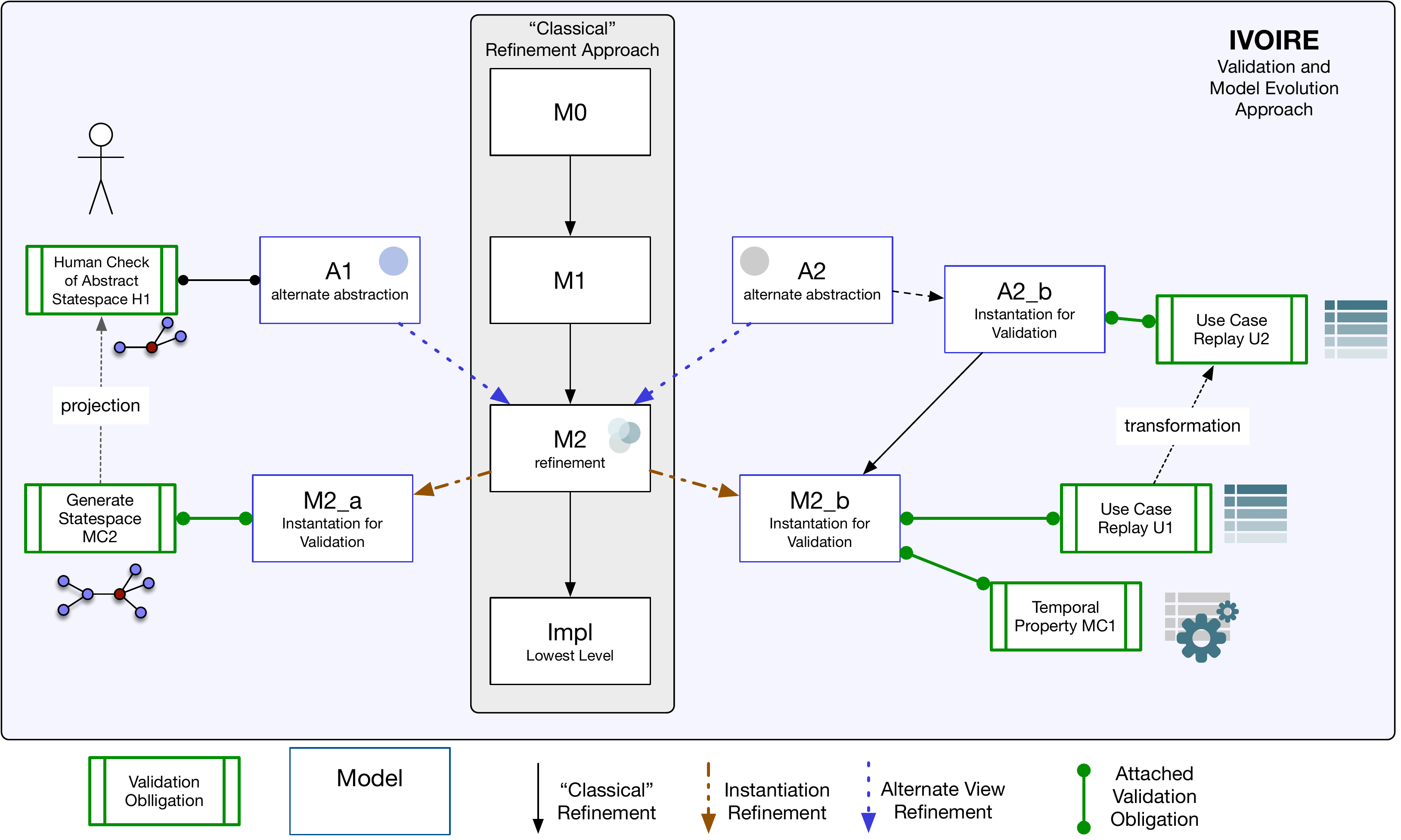}
    \caption{Refinement-based Software Development Process with VOs}
    \label{fig:software-process-vos}
\end{figure}
\end{center}

The idea of a refinement-based software development process assumes that a formal model is developed incrementally, i.e., step-by-step (see \cref{fig:software-process-vos}).
This means, that a model's refinement is created for each development step (black/solid-line arrows).
Later down the refinement chain, more requirements are taken into account.
VOs shall be used to ensure the presence of requirements in a refinement-based software development process.

To achieve this, newly introduced requirements must be validated incrementally at each model's refinement. 
Additionally, there might be the need for abstracting (illustrated by the blue/dotted arrows) or specializing/instantiating (illustrated by the red/dotted arrows) the model for a domain expert, only focusing on specific requirements.
Note that the concept of refinement is already supported in some formalisms (e.g. B and Event-B), but the concept of multiple distinct abstractions is novel, as far as we are aware.

\subsection{Refinement and Refactoring}

Refinement is an essential technique to enrich models while ensuring their correctness. 
As shown in \cref{fig:software-process-vos}, the refinement chain has a crucial role in the context of VOs, too. 
First, there is the classical refinement chain making up the middle of the figure. 
Here, the model is consecutively enriched with behavior and details. 
But there are also instantiation refinements and alternate view refinements going to the left and the right.

\textbf{Instantiation Refinements} are those that make a model particular for a use case, e.g., by providing an initialization for the variables.

\textbf{Alternate View Refinements} are those that allow a more abstract view onto the model, enabling easier reasoning, e.g., by ignoring behavior that is not relevant to validate a property, showing this property can become easier.\\
Multiple problems arise from that:
\begin{enumerate}
    \item How should such refinements interact with each other? On the right-hand side of \Cref{fig:software-process-vos}, one can see that $A2\_b$ is refined by $M2\_b$, which is also an instance of $M2$. It is an ongoing question of how the relationship between these components should be allowed and formally defined in the first place. The problem of these multi-layer relationships is keeping track of the changes and dependencies. Furthermore, every abstraction has to be validated, and in the case of the multi-layer relationship, the abstraction $A2$ has to be verified as the refinement of the instance of $A2\_b$. Even with the minimal example, this would result in a set of new proof obligations that would have to be shown to ensure a correct refinement in the first place. 
    \item How should VOs be refined? When doing linear refinement, a VO that holds on $M1$ has to hold on $M2$. But what about nonlinear refinement? A VO on $A2\_b$ has to hold too when going down the refinement chain, but how is this shown? Imagine introducing $M3$ refining $M2$. Does one needs to create an $A3\_b$ to show the VO?. There could be a case where this abstraction is no longer feasible as new behavior entangles components that were only loosely connected before. An idea would be to reduce every VO from a nonlinear refinement back to the origin in the linear refinement chain, which will be researched and discussed in the future. 
    \item On the right-hand side of \Cref{fig:software-process-vos} one can see a VO transformation. We do not know yet what this means in practice. And what the applications and restrictions are.
\end{enumerate}

Besides, formal refinement models can be refactored e.g. changing the name of variables, or altering the state space by changing the behavior of operations. In this case, VOs are validated similar to POs. There might be tool support to adapt the VOs and especially their tasks to this in the future. For now, this is not an immediate concern as refactoring should not change the behavior of a model but the quality of life of the modeler.

\subsection{Definition of Validation Tasks} 
Before we define the term VO, we will first formalize the subsidiary concept of a \emph{validation task} (VT).
A validation task (VT) is identified with an identifier,
and consists of a validation technique that is applied with the given
validation parameters to the corresponding context.
Executing a VT possibly modifies the internal state of the validation tool, e.g., consisting of the currently explored state space, and the current trace.
The notation we will use for a VT is as follows:

\begin{center}
{\small
$\mathtt{VT_{id}}$/$\mathtt{VT_{context}}$/$\mathtt{VT_{technique}}$: $\mathtt{VT_{parameters}}$
}
\end{center}

\subsection{Classification of Validation Techniques}

In the following, we explain various validation techniques, and how they are formalized as validation tasks.
There are various validation techniques one can use to validate a requirement.
\cref{table:vts} contains an overview of VTs we have assembled while conducting or inspecting a variety of case studies.
These case studies range from academic case studies (e.g., the ABZ landing gear
case study \cite{landinggear}, and the ABZ automotive case study \cite{LeuschelMutz}) to industrial applications in the railway industry (e.g., \cite{DBLP:conf/asm/HansenL0KKNNS18,DBLP:conf/rssrail/ComptierLMPM19}).
Regarding the future, \cref{table:vts} and operators might gradually evolve for new case studies or applications.
When explaining each validation technique, we refer to the name column in \cref{table:vts}.
Overall, VTs in \cref{table:vts} are expressed at a high level to keep the table generic.
The VO approach intends to work independently of the used formalism and tool.

\paragraph{Validation by Animation, Trace Replay, Testing}
Animation makes it possible for a human to execute the model interactively.
Some animators explore all transitions from the current state to the succeeding states. This is done by interpreting the
operational semantics of the used formalism on the model with all possible values for parameters and variables that are assigned
non-deterministically. Regarding the notion for a transition, possible means that the corresponding guard is met~\cite{role_validation,mashkoor17a,mashkoor17b}.

The main advantage of animation is that the user can interact with the model and view the model's state after executing an action.
Thus, this validation technique makes it possible to reason about the model more easily. When an animator explores all
succeeding transitions, the user also gets the information on which actions can be applied outgoing from the current state.
This eases the interaction with the model in a way that the user does not need to think about which input parameters
are required to constraint the guard. Nonetheless, it is then necessary to iterate over the possible values for parameters and
non-deterministically assigned variables which leads to a combinatorial explosion of possible transitions. \cite{role_validation}

Outgoing from an animation process, the modeler could store the resulting trace representing a scenario with certain behaviors.
Later on, the trace can be used to re-play the scenario, i.e., to check whether the scenario is still re-playable from the model (realized by \textbf{TR}).
Trace replay is applied similar to animation, but with the main difference that it is done automatically.

A trace $T$ consists of a list of the transitions $t_1, \ldots, t_n$. 
For each transition, the modeler could optionally add a predicate $\psi$ to be checked after re-playing the transition.
Here, we will use the notation $t$ $<$$\psi$$>$ for a transition $t$ and a predicate $\psi$.
If there is no postcondition, we will use the notation $t$ only.

Traces can then also be viewed as acceptance and unit tests which are well-known in traditional programming practice.
Thus, they can then be used to ensure the presence of certain behaviors in the model.

Note that the form of a transition depends on the used formalism.
E.g., in the B method, a transition consists of the operation's name, the values for parameters, and the values for non-deterministic assigned variables.

\paragraph{Validation by Simulation}


Simulation such as \emph{co-simulation} \cite{DBLP:journals/simpra/ThuleLGML19} or \emph{timed probabilistic simulation} \cite{vu2021validation} can be used to execute a model automatically. Here, the modeler can define respective configurations or annotations to define simulation scenarios.
Monte Carlo simulation \cite{montecarlo} can be applied in the context of timed probabilistic simulation to generate a various number of simulations.
Based on the resulting execution runs, statistical validation techniques such as \emph{hypothesis testing} \cite{hypothesistest} (realized by \textbf{HT}), or \emph{estimation of probability} \cite{fisher_1925} (realized by \textbf{EOP}) can be applied to show the presence of a behavior \cite{vu2021validation}.

Sometimes, systems consist of several different components or subsystems that interact with each other.
Each subsystem might be embedded into a different tool, or even modeled with a different formalism.
Co-simulation implements the idea of combining the components into an overall system for simulation.
In particular, the subsystems and their communication with each other are simulated in parallel.
Regarding the communication itself, subsystems must exchange data with each other which again might trigger events. \cite{DBLP:journals/simpra/ThuleLGML19}

Regarding \emph{timed probabilistic simulation}, the modeler can simulate the underlying model with timing and probabilistic behavior.
Each simulation results in a trace where each executed event is annotated with a certain time, called \emph{timed trace}.
A timed trace can then be replayed in real-time, i.e., wall-clock time.

\paragraph{Validation by Test Case Generation}

Test case generation tries to satisfy a given coverage criterion by generating tests for a model.
The desired coverage criterion is satisfied if each possible branch is covered by a test.
Thus, each generated test is represented by a trace which can be seen as a scenario representing a certain property.
Therefore, test case generation is a validation technique that can be used to generate new scenarios which
again can be validated by animation, trace replay, and testing.
Coverage criteria include operation coverage (realized by \textbf{OC}) and MC/DC coverage (realized by \textbf{MCDC}).
While the goal of operation coverage is to cover each operation, MC/DC coverage is used to cover all possible outcomes of each operation~\cite{bonfanti20a,testcase_1,testcase_2}.

\paragraph{Validation by Model Checking}

Explicit-state model checking checks state-based behaviors of a system by exploring its state space exhaustively (realized by \textbf{MC}).
Exhaustive exploration leads to full coverage of the system's behavior when the model checking process terminates.
Furthermore, it is then ensured whether the property is fulfilled or not. In the case that a property is violated,
explicit-state model checking can return a counter-example.
Again, when applying model checking to find a state satisfying a certain property, the technique can also provide a path leading to this state.
Nevertheless, explicit-state model checking often struggles with the combinatorial explosion of the state space which is called the
state space explosion problem.  This is because the number of states in a state space grows exponentially wrt. the
number of variables in a model. \cite{modelchecking}

Temporal model checking includes LTL and CTL model checking.
LTL model checking checks a temporal property (expressed as LTL formula) that is expected for the given system (realized by \textbf{LTL}).
Using the transition system and the Büchi automaton that is created from the LTL formula, 
LTL model checking checks temporal properties which are more complex than state-based properties. 
When negating the LTL formula, one is also able to find an example where the temporal property is true. \cite{modelchecking}

To formulate more expressive temporal properties, the modeler could also write CTL formulas and apply CTL model checking (realized by \textbf{CTL}).
Compared to LTL, CTL supports the operators $A \phi$ ($\phi$ is true for all paths), and $E \phi$ (it exists at least one path where $\phi$ is true).
\cite{modelchecking}

As the state space is also explored exhaustively, there are the same advantages and disadvantages as explicit-state model checking. \cite{modelchecking}

Symbolic model checking (realized by \textbf{SMC}) bases on the idea of getting rid of the state-space explosion problem.
To achieve this, the state space is not explored explicitly. 
Instead, logical formulae are derived from the model and then checked for solutions where properties are violated.  
Symbolic model checking makes use of techniques such as
SMT solving and abstract interpretation which are realized in the algorithms for constraint-based model checking,
bounded model checking, k-Induction and IC3 etc.
As the symbolic evaluation of the model is an over-approximation, there might be some false positives. Furthermore, the counter-example might also be abstracted which leads to a loss of information. \cite{krings2017towards}

By assigning probabilities to events in a model, a state space could be generated on which transitions are labeled with probabilities.
As result, the state space can be viewed as a Markov chain on which probabilistic model checking can be applied to validate probabilistic properties.
It is also possible to validate probabilistic temporal properties, e.g., properties that are encoded with PLTL, PCTL, or PB-LTL formulas.
Similar to probabilistic model checking, statistical model checking also aims to check probabilistic properties.
The main difference is that statistical model checking applies Monte Carlo simulation, whereupon PB-LTL or BLTL formulas
are checked with hypothesis testing or estimation. \cite{smc1, smc2}
Both model checking techniques are realized by \textbf{PSMC}.

As mentioned before, timed probabilistic simulation also applies Monte Carlo simulation together with statistical validation techniques.
However, timed probabilistic simulation does not check temporal formulas.
Instead, the modeler can specify a property (with timing behavior if desired) along with a start and end condition which should be checked. \cite{vu2021validation}

\paragraph{Validation by Proving}

Proving is a technique that is used to ensure the model's consistency, i.e., to show
the correctness of the program in certain aspects.
To achieve this, proving is often applied to proof obligations which are formulas that are generated
from the model (realized by \textbf{PO}).
Relevant aspects could be e.g., the violation of invariants, deadlocks, well-definedness errors, or
refinement errors.
The process of proving itself is both automatic and interactive  \cite{abrial2007rodin}.

In practice, different solvers are applied to try to prove a formula.
However, solvers are sometimes not strong enough to prove a formula.
The proof must then be done by the user interactively with additional effort.

The main purpose of proving is to ensure that the model does not contain any errors which seems to be rather verification than validation.
As discussed in \cref{sec:vandv}, we also see proving as validation.

\paragraph{Validation by Visualization, Statistics, and Metrics}

Another category of validation techniques includes inspection of \emph{visualizations}, \emph{tables}, \emph{statistics}, and \emph{metrics}.
The following VTs we consider are:

\begin{itemize}

\item \emph{State Space Visualization:} One of such techniques includes visualizing and inspecting the whole state space after applying certain steps. For a given state space consisting of reachable states, and possible transitions, one can formulate a predicate over the state space to be checked. This is realized in \textbf{SVIS}.

\item \emph{State Space Projection:} In practice, state spaces often become very large due to the state space explosion problem.
As result, the visualization gets too complex to understand.
To solve this problem, the modeler could provide an expression to create a state space projection onto this expression.
This results in an abstract visualization of the state space which is easier to understand. \cite{DBLP:conf/icfem/LadenbergerL15}
Similar to the state space inspection, one could also formulate a predicate to be checked over a projected state space (realized by \textbf{SPRJ}).
The projected state space could also provide a base to apply other VTs, e.g., LTL model checking.

\item \emph{Enabling Diagram}:
An enabling diagram is a diagram that describes for each operation which operations is enabled after executing this operation \cite{enabling}.
This helps to inspect how operations could depend on each other.
The corresponding VT (see \textbf{ED}) expects a formula that is checked over the diagram.

\item \emph{Operation Coverage Table}: This table describes for each operation whether it is covered yet.
Formally, one can provide a formula that is checked on the table as shown in \textbf{OCT}.
Combining this task with other VTs, one can evaluate the coverage of the other VTs.
For example, \textbf{OCT} could be applied after running a set of scenarios.
Afterwards, one can then measure the coverage, and thus also the quality of the given set of scenarios.

\item \emph{Read/Write Matrix}:  
The read/write matrix is a table describing for each operation which variables are read and written.
Similarly, one can inspect this matrix by checking a certain predicate (realized in \textbf{RWM}).
This helps to inspect which parts of a state are influenced by an operation and vice versa.

\item \emph{Variable Coverage Table}: This table provides the number of values a variable has been assigned to. 
Similarly, one can also provide a predicate that is checked over the table (realized in \textbf{VCT}).
This task is usually also combined with other VTs, to evaluate their coverage wrt. to the model's variables.

\item \emph{Min/Max Values for Variables}: This table shows the minimum and maximum value each variable has been assigned to.
Based on the table, one can also formulate a predicate which is checked on the table as shown in \textbf{MMV}.
In combination with other VTs, this task also helps to inspect the coverage.
For example, one can then inspect why the state space explodes.

\item \emph{State Space Statistics}:
To validate a model, the modeler could also take state space statistics into account.
Interesting statistics for the state space, could be, e.g., the number of states, or the number of transitions.
One could also extract more complex statistics, e.g., the number of states with a certain property such as
invariant violation, deadlock, or liveness.
To determine which events are particularly important for the model, one could also take the number of transitions
for each event into account.
This task is also formalized with a predicate over the state space statistics as shown in \textbf{STAT}.

\item \emph{Simulation Statistics}:
Based on a model and a simulation, one can also inspect simulation statistics.
Interesting properties are, e.g., statistics about how frequently an operation is executed in the simulation.
One could also inspect the percentage of how frequently an operation is executed when it is enabled.
Similar to \textbf{STAT}, this is also formalized with a predicate over the statistics (see \textbf{SISTAT}).

\end{itemize}

\paragraph{Vacuous Guards/Invariants:} There are also requirements made about the model's internal structure, rather than its functionality. For example, there could be a requirement, desiring that there are no vacuous parts in the invariant or an operation's guard.
Therefore, we have introduced the corresponding VT which is formalized as \textbf{VAP}.

\paragraph{Example:}

For example, the following VT identified by $\mathbf{LTL_{1}}$ means that LTL model checking should be applied with the LTL formula \texttt{G\{tl\_peds = red $\vee$ tl\_cars = red\}} expecting a successful result in the Traffic Light model.

\begin{framed}
{\scriptsize
$\mathbf{LTL_{1}}$/TrafficLight/LTL: G\{tl\_peds = red $\vee$ tl\_cars = red\}, SUCCESS
}
\end{framed}

\begin{table}[htbp]
\caption{Classification of Validation Tasks}
\label{table:vts}
\resizebox{\textwidth}{!}{
\begin{tabular}{lllll}
\toprule
Name & Task & Context & Parameters & Discharged \\ \midrule
TR $^4$ $^5$ & Trace Replay/ & Model & Trace $T$ & Automatic$^1$ \\
& Animation & & & \\ \addlinespace
HT $^4$ $^6$ & Simulation + & Model, & Hypothesis $H$ $^3$, Significance level $\alpha$  & Automatic \\
& Hypothesis Testing & Simulation & & \\
EOP $^4$ $^6$ & Simulation + & Model, & Property $P$ $^3$, Delta value $\delta$  & Automatic \\
& Estimation of Probability & Simulation & & \\ \addlinespace
OC $^4$ $^6$ & Operation Coverage & Model & Operations $O$ & Automatic \\
& Test Case Generation & & & \\
MCDC $^4$ $^6$ & MC/DC Test Case & Model & Level $l$ & Automatic \\
& Generation & & & \\ \addlinespace
MC $^4$ $^5$ $^6$ & Explicit-state & Model & Configuration $c \in \mathit{Conf_{MC, F}}$ & Automatic \\
& Model Checking & & & \\
SMC $^6$ & Symbolic Model Checking & Model & Configuration $c \in \mathit{Conf_{MC, F}}$ & Automatic \\
LTL $^4$ $^5$ $^6$ & LTL Model Checking & Model & LTL Formula $\psi$, $c \in \{\mathtt{SUCCESS}, \mathtt{FAIL}\}$ & Automatic$^2$ \\
CTL $^4$ $^5$ $^6$ & CTL Model Checking & Model & CTL Formula $\psi$, $c \in \{\mathtt{SUCCESS}, \mathtt{FAIL}\}$ & Automatic$^2$ \\
PSMC $^4$ $^6$ & Probabilistic/Statistical & Model & Probabilistic Temporal Formula $\psi$ $^3$, & Automatic$^2$ \\
& Model Checking & & $c \in \{\mathtt{SUCCESS}, \mathtt{FAIL}\}$ & \\ \addlinespace
PO & Proving & Model & Formula $\psi$ & Partial \\ \addlinespace
SVIS $^7$ & Inspection of & Model & Formula $\psi$ over state space& Manual $^9$ \\
& State Space Visualization & & $S_\mathtt{vis} = (Z_\mathtt{vis}, T_\mathtt{vis})$ & \\
SPRJ $^7$ & Inspection of & Model & Expression $\phi$, Formula $\psi$ over projected & Manual $^9$ \\
& State Space Projection & & state space on $\phi$: $S_{\phi} = (Z_{\phi}, T_{\phi})$ & \\ \addlinespace
STAT $^8$ & Inspection of & Model & Formula $\psi$ over state space statistics  & Manual $^9$ \\
& State Space Statistics & & $R_\mathtt{stat} \subseteq P_\mathtt{stat} \times \mathbb{R}_{\geq 0}$ $^3$ $^{10}$ & \\
SISTAT $^8$ & Simulation +  & Model, & Number of Simulations $N$, & Manual $^9$ \\
& Inspection of & Simulation & Start Condition $C_\mathtt{st}$, End Condition $C_\mathtt{end}$, & \\
& Simulation Statistics & & Formula $\psi$ over simulation statistics \\ 
& & & $R_\mathtt{sistat} \subseteq P_\mathtt{sistat} \times \mathbb{R}_{\geq 0}$ $^3$ $^{10}$ & \\
ED $^7$ & Inspection of & Model & Formula $\psi$ over enabling diagram & Manual $^9$ \\ 
& Enabling Diagram & & $R_\mathtt{ed} \subseteq E \times E$ $^{11}$ & \\
OCT $^8$ & Inspection of & Model & Formula $\psi$ over operation coverage table & Manual $^9$ \\
& Operation Coverage Table & & $R_\mathtt{oct} \subseteq E \times \{\mathtt{COV}, \mathtt{UNCOV}\}$ $^{11}$ & \\
RWM $^8$ & Inspection of & Model & Formula $\psi$ over read/write matrix & Manual $^9$ \\
& Read/Write Matrix & & $R_\mathtt{rwm} \subseteq \{\mathtt{READ}, \mathtt{WRITE}\} \times (E \times V)$ $^{11}$ & \\
VCT $^8$ & Inspection of & Model & Formula $\psi$ over variable coverage table & Manual $^9$ \\
& Variable Coverage Table & & $R_\mathtt{vct} \subseteq V \times \mathbb{N}_{0}$ $^{11}$ & \\
MMV $^8$ & Inspection of & Model & Formula $\psi$ over min/max values table & Manual $^9$ \\
& Min/Max Values & & $R_\mathtt{mmv} \subseteq V \times (\bigcup_{t \in V} \tau(t) \times \bigcup_{t \in V} \tau(t))$ $^{11}$ $^{12}$ & \\ \addlinespace
VAP & Vacuous Parts Check & Model & Configuration $c \in \{\mathtt{GRD}, \mathtt{INV}\}$ & Automatic \\
\bottomrule
\end{tabular}}
\newline
\scriptsize{$^1$ succeeds when scenario is replayable and all tests succeed\newline
$^2$ succeeds when $\psi$ is fulfilled for SUCCESS or $\psi$ failed for FAIL\\
$^3$ depends on the tool that is used \\
$^4$ explores state space \\
$^5$ updates current trace (model checking only for FAIL and GOAL) \\
$^6$ generates (counter)-examples for a domain expert \\
$^7$ generates visualization for a domain expert \\
$^8$ generates table for a domain expert \\
$^9$ succeeds when formula is fulfilled on visualization, table, statistics, ... \\
$^{10}$ $P_{stat}$ and $P_{sistat}$ denote the respective set of (simulation) statistics properties \\
$^{11}$ $E$ denotes the set of events/operations, $V$ denotes the set of variables \\
$^{12}$ $\tau(v)$ denotes the set of all values which can be of $v$'s type \\

Remark: $\mathit{Conf_{MC, F}}$ is currently defined as $\{<\mathtt{FIN}>, <\mathtt{DLF}>\} \cup \{<\mathtt{INV}, \psi> \mid \psi \in F\} \cup \newline \{<\mathtt{GOAL}, \psi> \mid \psi \in F\}$, where $F$ denotes the set of formulas in the supported formalism, FIN denotes checking for finite state-space, DLF denotes checking for deadlock-freedom, INV denotes invariant checking, and GOAL denotes searching for a goal.}

\end{table}

\paragraph{Code Generation for Validation}
In contrast to the aforementioned techniques, we do not see code generation as a validation technique directly.
Instead, it is rather a tool that can be applied to enable other validation techniques afterwards.

During the software development process using formal methods, software is specified and refined step by step.
Once a refinement level is reached which is close to implementation constructs, a code generator is applied.
Regarding the B method, a code generator for embedded systems can be applied once the B0 language is reached, which is the implementable subset of B \cite{clearsy2016atelierb}.
As an implementable subset of the specification language is required, memory usage of the final refinement can be verified.
Thus, the generated code can be used for embedded systems.
Additionally, the software engineer could also write or generate tests to validate the generated code.
This also means that these validations are applied at the very end of the software development process.

Communication with the stakeholder and early-stage validation is particularly important in the context of VOs.
To achieve this goal, our approach intends to take high-level code generators such as \btoprogram{} \cite{vu2019multi} into account.
\btoprogram{} is suitable for application for early-stage validation of the software, but cannot be used to generate code for embedded systems.
Based on the generated code, the model could then be animated, tested, and simulated.
As the model is translated to a programming language, it could also be more familiar to the domain expert to work with it compared
to working in the context of formal methods.

\paragraph{Languages and their tools}
For our research we have investigated nine major modeling languages regarding their tool support for different tasks. The results are shown in \cref{tools_abilities}. Whenever something is  marked with \xmark,  we did not find referable evidence for the existence of the respective tool support. A more comprehensive evaluation of state-based formal methods is provided by Mashkoor et al.~\cite{mashkoor18a}.\\
One can see that tool support is widely spread. As we use the ProB platform as starting point for further development, the B and Event-B languages are especially appealing as they are covered by most of the features we investigated.

\begin{table}[htbp]
\centering
\caption{Specification Languages and Supported Validation Techniques}
\label{tools_abilities}
\resizebox{\textwidth}{!}{
\setlength\tabcolsep{1pt}

\begin{tabular}{l|lllllllll}
Tools & Alloy & ASM & B & Event-B & VDM & TLA+ & Z & CSP & Circus \\ \hline
Animation & \xmark & \cmark(\cite{asmetaa})& \cmark(\cite{leuschel2003prob, visB})& \cmark(\cite{prob2, leuschel2003prob, visB})& \cmark(\cite{oda2015vdm}) & \cmark(\cite{hansen2012translating}) & \cmark(\cite{prob2, derrick2008z2sal, plagge2007validating}) & \cmark(\cite{prob2, ProBCSP}) & \xmark \\
Trace Replay/Testing & \cmark(\cite{alloyAnalyzer, electrum}) \tablefootnote{In Alloy, it seems that it is not possible to animate the model interactively. Nonetheless, it is still possible to test the feasibility and behavior of a scenario. Here, it seems that scenarios have to be encoded manually. Furthermore, note that Alloy only supports infinite traces} & \cmark(\cite{asmetav}) & \cmark(\cite{prob2_ui}) & \cmark(\cite{snook2021domain, prob2_ui})& \xmark & \cmark(\cite{hansen2012translating, prob2_ui}) & \cmark(\cite{z_replay}) & \cmark(\cite{carter2007mise}) & \xmark \\ \midrule
Test Case Generation & \cmark(\cite{sullivan2017automated}) & \cmark(\cite{Gargantini2001ASMBasedTC,gargantini2007using,gargantini2003using}) & \cmark(\cite{prob_test_case_generation}) & \cmark(\cite{testcase_1,testcase_2, prob_test_case_generation}) & \cmark(\cite{droschl1999design}) & \cmark(\cite{hansen2012translating, prob_test_case_generation}) & \cmark(\cite{helke1997automating, prob_test_case_generation}) & \cmark(\cite{testGenerationCSP}) & \xmark \\ \midrule
Simulation & \cmark(\cite{brunel2019simulation}) & \cmark(\cite{asmetas}) & \cmark(\cite{vu2021validation}) & \cmark(\cite{vu2021validation}) & \cmark(\cite{DBLP:journals/simpra/ThuleLGML19,CoSimulationVDM}) & \cmark(\cite{vu2021validation}) & \cmark(\cite{vu2021validation}) & \cmark(\cite{vu2021validation}) & \questionmark \\ \midrule

Explicit-State MC & \cmark(\cite{brunel2018electrum, electrum}) & \cmark(\cite{asmetamc}) & \cmark(\cite{leuschel2003prob}) & \cmark(\cite{prob2, leuschel2003prob})& \cmark(\cite{lin2015towards}) & \cmark(\cite{yu1999model, hansen2012translating})& \cmark(\cite{plagge2007validating, prob2}) & \cmark(\cite{gibson2014fdr3, prob2}) & \xmark \\
LTL MC & \cmark(\cite{brunel2018electrum, electrum}) & \cmark(\cite{asmetamc})& \cmark(\cite{plagge2010seven}) & \cmark(\cite{plagge2010seven})& \cmark(\cite{lin2015towards}) & \cmark(\cite{hansen2012translating, plagge2010seven}) & \cmark(\cite{plagge2010seven, derrick2011z2sal})& \cmark(\cite{plagge2010seven, csp_spin}) & \xmark \\
CTL MC & \cmark(\cite{vakili2012temporal}) & \cmark(\cite{asmetamc})& \cmark(\cite{leuschel2003prob})& \cmark(\cite{prob2, leuschel2003prob})& \cmark(\cite{lin2015towards}) & \cmark(\cite{prob2, leuschel2003prob, hansen2012translating}) & \cmark(\cite{derrick2011z2sal}) & \cmark(\cite{leuschel2003prob}) & \xmark \\
Symbolic MC & \cmark(\cite{brunel2018electrum, electrum}) & \cmark(\cite{asmetamc})& \cmark(\cite{krings2017towards}) & \cmark(\cite{krings2017towards}) & \xmark & \cmark(\cite{konnov2019tla+, konnov2018bmcmt, krings2017towards}) & \cmark(\cite{symbolicZ}) & \cmark(\cite{sun2008bounded}) & \xmark \\
Probabilistic/Statistical MC & \xmark & \xmark & \xmark & \cmark (\cite{event-b-probabilistic-1, event-b-probabilistic-2}) & \xmark & \xmark & \xmark & \xmark & \xmark \\ \midrule
Proving & \cmark(\cite{ulbrich2011proving}) & \cmark(\cite{kiv}) & \cmark(\cite{mentre2012discharging}) & \cmark(\cite{abrial2005b, abrial2007rodin}) & \cmark(\cite{agerholm1996translating}) & \cmark(\cite{chaudhuri2008tla}) & \cmark(\cite{zproof, Freitas2005ProvingTW}) & \cmark(\cite{cspSolving, isobe2008proof}) & \cmark(\cite{freitas2005model}) \\ \midrule
Refinement Checking & \xmark & \cmark(\cite{arcaini2016smt, borger2003asm}) & \cmark(\cite{mentre2012discharging}) & \cmark(\cite{abrial2005b, abrial2007rodin}) & \cmark(\cite{maharaj1997verification}) & \cmark(\cite{taibi2009stepwise}) & \cmark(\cite{stringer1997using})& \cmark(\cite{gibson2014fdr3})& \cmark(\cite{freitas2005model}) \\ \midrule
State Space Visualization & \questionmark & \questionmark & \cmark(\cite{DBLP:conf/icfem/LadenbergerL15}) & \cmark(\cite{DBLP:conf/icfem/LadenbergerL15}) & \xmark & \cmark(\cite{Kuppe_2019, DBLP:conf/icfem/LadenbergerL15}) & \cmark(\cite{DBLP:conf/icfem/LadenbergerL15}) & \cmark(\cite{leuschel2003prob, prob2, ProBCSP}) & \xmark \\ \midrule 

Code Generation \tablefootnote{High-Level Code Generation for (Early-Stage) Validation} & \xmark & \cmark(\cite{bonfanti2017asm2c++})& \cmark(\cite{vu2019multi})& \cmark(\cite{rivera2017code}) & \cmark(\cite{hasanagic2019code}) & \xmark & \xmark & \cmark & \cmark (\cite{barrocas2012jcircus}) \\ \bottomrule


\end{tabular}
}

\end{table}

\subsection{Definition of Validation Obligation}

We define the term \emph{validation obligation} as follows:

\begin{quote}
A \emph{validation obligation} (VO) is composed of (multiple) validation tasks (VT) associated with a model to check its compliance with the requirement.
\end{quote}

Thus, validating a requirement succeeds if all associated VOs yield successful results.
Formally, a VO is annotated with an id, and consists of a \emph{validation expression} (VE).
The VE consists of operations on the associated VTs.
Therefore, a VO succeeds, if the corresponding $\mathrm{VO_{expression}}$ leads to a successful result.
The notation we will use for a VO is as follows:

\begin{center}
{\small
$\mathtt{VO_{id}}: \mathtt{VO_{expression}}$
}
\end{center}


An example for a VO that validates $\mathbf{LTL_{1}}$ is shown in $\mathbf{VO_{1}}$.
$\mathbf{VO_{1}}$ succeeds if $\mathbf{LTL_{1}}$ discharges successfully.

\begin{center}
$\mathtt{VO_{1}}$: validate($\mathtt{LTL_{1}}$)
\end{center}

A VE with a single VT $\mathtt{T}$ succeeds if $\mathtt{T}$ succeeds.
For a VE $\mathtt{T}$, we allow the unary operator $\neg \;$ $\mathtt{T}$ which succeeds if validating $\mathtt{T}$ fails. 
For two VEs $\mathtt{T_{1}}$ and $\mathtt{T_{2}}$, we allow the following logical operators: $\mathtt{T_{1}}$ $\wedge$ $\mathtt{T_{2}}$ ($\mathtt{T_{1}}$ and $\mathtt{T_{2}}$ must succeed), $\mathtt{T_{1}}$ $\vee$ $\mathtt{T_{2}}$ ($\mathtt{T_{1}}$ or $\mathtt{T_{2}}$ must succeed), $\mathtt{T_{1}}$ $\Rightarrow$ $\mathtt{T_{2}}$ (if $\mathtt{T_{1}}$ succeeds then $\mathtt{T_{2}}$ must succeed as well), $\mathtt{T_{1}}$ $\Leftrightarrow$ $\mathtt{T_{2}}$ ($\mathtt{T_{1}}$ must succeed when $\mathtt{T_{2}}$ succeeds and vice versa).
Additionally, we also consider the sequential composition $\mathtt{T_{1}} ; \mathtt{T_{2}}$ which means that $\mathtt{T_{2}}$ is executed after $\mathtt{T_{1}}$, based on $\mathtt{T_{1}}$'s result.
Thus, $\mathtt{T_{1}} ; \mathtt{T_{2}}$ succeeds, if $\mathtt{T_{1}}$ succeeds, and $\mathtt{T_{2}}$ succeeds after executing $\mathtt{T_{1}}$.
For example, $\mathtt{T_{1}}$ could apply model checking searching for a state, which is then used as the initial state of a trace replay task $\mathtt{T_{2}}$.
Regarding the future, the set of operators might evolve.
We also implemented a semantic checker to check the VE's consistency.
For example, evaluating the coverage without applying any tasks before does not have a meaning and therefore is semantically wrong.

\subsection{Creating Validation Obligations}
Currently, a VO is created by the modeler manually.
There are also tools like UML-B \cite{snook2006uml}, which attempt an automatic translation from the specification to a model.
Regarding the future, one could explore whether and how VOs can be extracted from the requirements automatically.
Here, we could take \textsc{FRETIish} or \textsc{SpeAR} into account to write behavioral requirements in natural language.

When creating a VO, the modeler needs significant knowledge about the modeling language, and about the environment and the techniques to create suitable tasks.
For example, the preservation of an invariant can be shown by model checking or proving. 
Proving and symbolic model checking are therefore more suitable than explicit-state model checking to check an invariant in an infinite-state system.
Another aspect is to check whether the property formulated in the VO actually captures the stakeholders' needs.
As natural language is ambiguous, communication and feedback from the stakeholders are important.

\subsection{VO-guided Workflow}
Requirements engineering and software development are highly entangled processes.
During the software development process, requirements are encoded into the model incrementally.
When validating those requirements, stakeholders and developers get a better understanding of the system
which might lead to new requirements being discovered, or existing requirements evolving or being changed.
In the case that a VO fails, the modeler needs to re-consider the VO, the requirement, or even the model.
Possible questions that could be asked are:

\begin{itemize}
\item Did we translate the requirement into a VO, a VT task or the model poorly? 
\item Does the requirement collide with other requirements? 
As a result, we may need to weaken or strengthen this or other requirements.
\end{itemize}





Another important aspect is requirements engineering, to structure requirements systematically, and to define dependencies between them.
For example, there are languages such as \textsc{KAOS} \cite{kaos}, \textsc{DOORS} \cite{doors}, or Problem Frames \cite{problem_frames}.
Furthermore, there is also \textsc{ProR} which defines an approach to structure requirements \cite{Jastram2012ThePA}.
Concerning refinement and traceability, it will also be important to define dependencies between requirements and thus also VOs.
Therefore, we will also take those aforementioned works into account.

\section{Demonstration of Validation Obligations}
\label{sec:demonstration}

In this section, we will demonstrate how VOs can be used to validate requirements.

\textbf{Requirements} Let us consider a small traffic light example, modeling the cars' traffic light and the pedestrians' traffic light at a crossing in Germany, with the following requirements:

\small
\begin{framed} 
\textbf{FUN1:} There are two traffic lights: the cars' traffic light and the pedestrians' traffic light. Initially, both traffic lights are red.
\end{framed}
\normalsize

\small
\begin{framed} 
\textbf{FUN2:} Cars' traffic light can switch to red and yellow, if it is red and the pedestrians' traffic light is red.
\end{framed}
\normalsize

\small
\begin{framed} 
\textbf{FUN3:} Cars' traffic light can switch to green, if it is red and yellow and the pedestrians' traffic light is red.
\end{framed}
\normalsize

\small
\begin{framed} 
\textbf{FUN4:} Cars' traffic light can switch to yellow, if it is green and the pedestrians' traffic light is red.
\end{framed}
\normalsize

\small
\begin{framed} 
\textbf{FUN5:} Cars' traffic light can switch to red, if it is yellow and the pedestrians' traffic light is red.
\end{framed}
\normalsize

\small
\begin{framed} 
\textbf{FUN6:} Pedestrians' traffic light can switch to green, if it is red and the cars' traffic light is red.
\end{framed}
\normalsize

\small
\begin{framed} 
\textbf{FUN7:} Pedestrians' traffic light can switch to red, if it is green and the cars' traffic light is red.
\end{framed}
\normalsize

\small
\begin{framed} 
\textbf{SAF1:} One of both traffic lights is red at any moment.
\end{framed}
\normalsize

\small
\begin{framed}
\textbf{SAF2:} Cars' traffic light can either be red, red and yellow, yellow, or green.
\end{framed}
\normalsize

\small
\begin{framed} 
\textbf{SAF3:} Pedestrians' traffic light can either be red, or green.
\end{framed}
\normalsize

\small
\begin{framed} 
\textbf{LIV1:} The situation that both traffic lights are red occurs infinitely often. 
\end{framed}
\normalsize

\small
\begin{framed} 
\textbf{SCENARIO1: Running Cycle for Cars' Traffic Light:} \\
In the beginning, the cars' and the pedestrians' traffic light are both red. \\
The cars' traffic light then switches from red to red and yellow. \\
Afterwards, it switches from red and yellow to green. \\
Now, it switches back to yellow, and then to red. \\
The pedestrians' traffic light stays red during the scenario.
\end{framed}
\normalsize

\small
\begin{framed} 
\textbf{SCENARIO2: Running Cycle for Pedestrians' Traffic Light:} \\
In the beginning, the cars' and the pedestrians' traffic light are both red. \\
The pedestrians' traffic light switches from red to green. \\
Afterwards, it switches back from green to red. \\
The cars' traffic light stays red during the scenario.
\end{framed}
\normalsize

Furthermore, this report also considers additional functional requirements \textbf{FUN8}, \textbf{FUN9}, and \textbf{FUN10} in a refinement.
These requirements will not be validated in this report.

\small
\begin{framed} 
\textbf{FUN8:} A controller can send a command to switch a traffic light to a specific color, if there are no other commands queued.
\end{framed}
\normalsize

\small
\begin{framed} 
\textbf{FUN9:} A traffic light can only switch its color if there is a corresponding command queued.
\end{framed}
\normalsize

\small
\begin{framed} 
\textbf{FUN10:} A command can be rejected after it has been sent by the controller.
\end{framed}
\normalsize

Encoding those functional requirements leads to the B model described in \cref{lst:b-trafficlight}.

\begin{lstlisting}[
caption=Traffic Light Example,
label=lst:b-trafficlight]
MACHINE TrafficLight
SETS colors = {red, redyellow, yellow, green}
VARIABLES tl_cars, tl_peds

INVARIANT tl_cars : colors & tl_peds : {red, green} &
	     (tl_peds = red or tl_cars = red)
              
INITIALISATION  tl_cars := red || tl_peds := red
	
OPERATIONS
cars_ry = SELECT tl_cars = red & tl_peds = red THEN tl_cars := redyellow END;
cars_y = SELECT tl_cars = green THEN tl_cars := yellow END;
cars_g = SELECT tl_cars = redyellow THEN tl_cars := green END;
cars_r = SELECT tl_cars = yellow THEN tl_cars := red END;
peds_r = SELECT tl_peds = green THEN tl_peds := red END;
peds_g = SELECT tl_peds = red & tl_cars = red THEN tl_peds := green END
      
END
\end{lstlisting}

To demonstrate state space projection, \cref{lst:b-trafficlight} will be refined.
Regarding Classical B, it is not only necessary to add new events in the refinement, but also to add them in the abstract machine refining \texttt{skip}.
The resulting machines are shown in \cref{lst:b-trafficlight-command-abstract} and \cref{lst:b-trafficlight-command-concrete}.

\begin{center}
\begin{figure}[ht]
    \centering
    \includegraphics[scale=0.4]{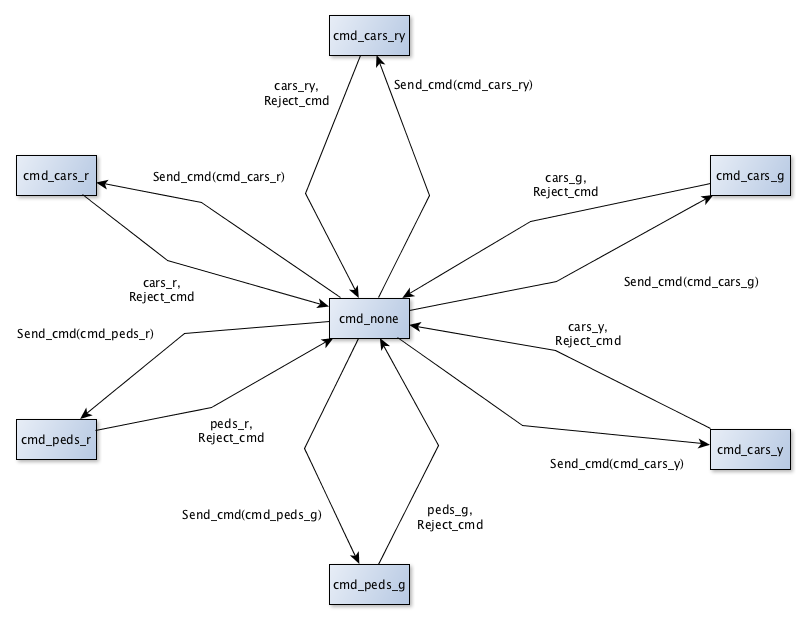}
    \caption{\textbf{PRC1} as Diagram}
    \label{projection-diagram}
\end{figure}
\end{center}

After encoding commands to switch the traffic lights' colors (\textbf{FUN8} - \textbf{FUN10}), a domain expert might be interested in sending commands without considering the traffic light's color only.
Here, the domain expert could define a diagram describing how the logic for sending commands has to work. 
This is shown in \textbf{PRC1} in \cref{projection-diagram}.

Based on the model, the designer could also run different simulations.

\begin{lstlisting}[
caption=Traffic Light Simulation (TrafficLight\_Sim),
label=lst:b-trafficlight-sim]
{
 "activations": [
  {"id":"$initialise_machine", "execute":"$initialise_machine",
   "activating":"choose"},
  {"id":"choose", "chooseActivation":{"cars_ry": "0.5", "peds_g": "0.5"}},
  {"id":"cars_ry", "execute":"cars_ry", "after":5000, "activating":"cars_g"},
  {"id":"cars_g", "execute":"cars_g", "after":500, "activating":"cars_y"},
  {"id":"cars_y", "execute":"cars_y", "after":5000, "activating":"cars_r"},
  {"id":"cars_r", "execute":"cars_r", "after":500, "activating":"choose"},
  {"id":"peds_g", "execute":"peds_g", "after":5000, "activating":"peds_r"},
  {"id":"peds_r", "execute":"peds_r", "after":5000, "activating":"choose"}
 ]
}
\end{lstlisting}

\Cref{lst:b-trafficlight-sim} shows a \simb{} file \cite{vu2021validation} that annotates operations with times and probabilities.
Within the first simulation shown in \cref{lst:b-trafficlight-sim},
the controller chooses between the cars' traffic light's cycle and the pedestrians' traffic light's cycle
with a probability of 50\% for each.
Whenever a traffic light turns green or red, it will not switch the color for 5 seconds.
Switching the cars' traffic light from red and yellow to green, and yellow to red always takes 500 ms.

Based on this simulation, the modeler could then validate the probabilistic timing requirements \textbf{PROB-TIM1} and \textbf{PROB-TIM2}.

\small
\begin{framed} 
\textbf{PROB-TIM1:} Whenever both traffic lights are red, the cars' traffic light will turn green with a probability of at least 80\% within the next 30 seconds.
\end{framed}
\normalsize

\small
\begin{framed} 
\textbf{PROB-TIM2:} Whenever both traffic lights are red, the pedestrians' traffic light will turn green with a probability of at least 90\% within the next 30 seconds.
\end{framed}
\normalsize

Based on the encoded model, the non-functional requirements are as follows:

After validating \textbf{SCENARIO1} and \textbf{SCENARIO2}, the following coverage criteria are expected to hold: \textbf{COV1}, \textbf{COV2}, and \textbf{COV3}.

\small
\begin{framed} 
\textbf{COV1:} Validating \textbf{SCENARIO1} and \textbf{SCENARIO2} covers the model such that the cars' traffic light switches between four colors, while the pedestrians' traffic light switches between two colors.
\end{framed}
\normalsize

\small
\begin{framed} 
\textbf{COV2:} Validating \textbf{SCENARIO1} and \textbf{SCENARIO2} covers all operations in the model.
\end{framed}
\normalsize

\small
\begin{framed} 
\textbf{COV3:} Validating \textbf{SCENARIO1} and \textbf{SCENARIO2} covers the whole state space consisting of six possible states (including root) and seven possible transitions.
\end{framed}
\normalsize

\small
\begin{framed} 
\textbf{STRUC1:} tl\_cars is written by cars\_ry, cars\_y, cars\_g, cars\_g only.
\end{framed}
\normalsize

\small
\begin{framed} 
\textbf{STRUC2:} tl\_peds is written by peds\_r and peds\_g only.
\end{framed}
\normalsize

\small
\begin{framed} 
\textbf{STRUC3:} There are no vacuous parts in the invariant and guards of the model.
\end{framed}
\normalsize

\small
\begin{framed} 
\textbf{STRUC4:} The operations enable each other as follows: cars\_ry enables cars\_g, cars\_g enables cars\_y, cars\_y enables cars\_r, cars\_r enables cars\_ry and peds\_g, peds\_g enables peds\_r, peds\_r enables peds\_g and cars\_ry.
\end{framed}
\normalsize

\small
\begin{framed} 
\textbf{STRUC5:} All operations are coverable in the model.
\end{framed}
\normalsize

\small
\begin{framed} 
\textbf{STRUC6:} MC/DC coverage with level 2 should be feasible in the model.
\end{framed}
\normalsize

\textbf{Validation by VOs}
Now, we will describe how all these requirements are validated by VOs.
Particularly, we will present at least one VO for each requirement.
Since the requirements described above do not necessarily have to be validated by VTs from all types, 
we will also present alternative VTs to demonstrate all VT types.
Here, we will mainly focus on validation in \prob{}.
Furthermore, the VOs are formalized using operators in the B method.
Regarding probabilistic model checking, we will also take an example in PRISM into account.




In order to validate \textbf{FUN1}, it is necessary to check whether both traffic lights are red in all initial states.
Thus, this requirement could be validated by a VO applying an LTL model check to expect a positive result, as shown in \textbf{VO1}.

\small
\begin{framed} 
\textbf{LTL1}/TrafficLight/LTL: \{tl\_cars = red $\wedge$ tl\_peds = red\}, SUCCESS
\end{framed}
\normalsize

\small
\begin{framed} 
\textbf{VO1}: LTL1
\end{framed}
\normalsize


For validation of \textbf{FUN2}, one needs to check that whenever the cars' traffic light is red and yellow, it has been red and yellow since both traffic lights are red, one step ago. Thus, this behavior can be validated by a VO applying an LTL model check to expect a positive result as shown in \textbf{VO2}.

\small
\begin{framed} 
\textbf{LTL2}/TrafficLight/LTL: G (\{tl\_cars=redyellow\} $\implies$ (\{tl\_cars=redyellow\} S \{tl\_cars=red $\wedge$ tl\_peds=red\})), SUCCESS
\end{framed}
\normalsize

\small
\begin{framed} 
\textbf{VO2}: LTL2
\end{framed}
\normalsize


For validation of \textbf{FUN3}, one needs to check that whenever the cars' traffic light is green, it has been green since the cars' traffic light is red and yellow, and the pedestrians' traffic light is red, one step ago. Thus, this behavior can be validated by \textbf{VO3} which applies an LTL model check \textbf{LTL3} to expect a positive result.

\small
\begin{framed} 
\textbf{LTL3}/TrafficLight/LTL: G (\{tl\_cars=green\} $\implies$ (\{tl\_cars = green\} S \{tl\_cars=redyellow $\wedge$ tl\_peds=red\})), SUCCESS
\end{framed}
\normalsize

\small
\begin{framed} 
\textbf{VO3}: LTL3
\end{framed}
\normalsize


For validation of \textbf{FUN4}, one needs to check that whenever the cars' traffic light is yellow, it has been yellow since the cars' traffic light is green, and the pedestrians' traffic light is red, one step ago. This behavior is also validated by a VO applying an LTL model check expecting a positive result.

\small
\begin{framed} 
\textbf{LTL4}/TrafficLight/LTL: G (\{tl\_cars=yellow\} $\implies$ (\{tl\_cars=yellow\} S \{tl\_cars=green $\wedge$ tl\_peds=red\})), SUCCESS
\end{framed}
\normalsize

\small
\begin{framed} 
\textbf{VO4}: LTL4
\end{framed}
\normalsize


For validation of \textbf{FUN5}, one needs to check two behaviors: 

\begin{itemize}
\item The cars' traffic light might change its color unequal red (realized by \textbf{LTL5.1.}).
\item Assuming that the cars' traffic light has already switched its color unequal to red: Whenever the cars' traffic light is red, it has been red since the cars' traffic light is yellow, and the pedestrians' traffic light is red, one step ago (realized by \textbf{LTL5.2.}).
\end{itemize}

While the first property is expected to fail, the second property is expected to hold.
Regarding the first behavior, it would also be possible to apply explicit-state model checking searching for a goal.
The validation is realized in \textbf{VO5}.

\small
\begin{framed} 
\textbf{LTL5.1}/TrafficLight/LTL: $\neg$ (F\{tl\_cars $\neq$ red\}), FAIL
\end{framed}
\normalsize

\small
\begin{framed} 
\textbf{LTL5.2}/TrafficLight/LTL: (\{tl\_cars = red\} W (\{tl\_cars $\neq$ red\} $\wedge$ G(\{tl\_cars=red\} $\implies$ (\{tl\_cars=red\} S \{tl\_cars=yellow $\wedge$ tl\_peds=red\})))), SUCCESS
\end{framed}
\normalsize

\small
\begin{framed} 
\textbf{VO5}: LTL5.1 $\wedge$ LTL5.2
\end{framed}
\normalsize


For validation of \textbf{FUN6}, one needs to check that whenever the pedestrians' traffic light is green, it has been green since the cars' traffic light is red, and the pedestrians' traffic light is red, one step ago. This behavior is validated by \textbf{VO6} applying an LTL model check \textbf{LTL6} expecting a positive result.

\small
\begin{framed} 
\textbf{LTL6}/TrafficLight/LTL: G (\{tl\_peds=green\} $\implies$ (\{tl\_peds=green\} S \{tl\_cars=red $\wedge$ tl\_peds=red\})), SUCCESS
\end{framed}
\normalsize

\small
\begin{framed} 
\textbf{VO6}: LTL6
\end{framed}
\normalsize


For validation of \textbf{FUN7}, one needs to check two behaviors: 

\begin{itemize}
\item The pedestrians' traffic light might change its color unequal red.
\item Assuming that the pedestrians' traffic light has already switched its color unequal to red: Whenever the pedestrians' traffic light is red, it has been red since the cars' traffic light is red, and the pedestrians' traffic light is green, one step ago.
\end{itemize}

Both behaviors can be formulated as an LTL model check, too. 
While the first property is expected to fail, the second property is expected to hold.
Regarding the first behavior, it would also be possible to apply explicit-state model checking searching for a goal.
The validation is realized in illustrated in \textbf{VO7}.

\small
\begin{framed} 
\textbf{LTL7.1}/TrafficLight/LTL: $\neg$ (F\{tl\_peds $\neq$ red\}), FAIL
\end{framed}
\normalsize

\small
\begin{framed} 
\textbf{LTL7.2}/TrafficLight/LTL: (\{tl\_peds = red\} W (\{tl\_peds $\neq$ red\} $\wedge$ G(\{tl\_peds=red\} $\implies$ (\{tl\_peds=red\} S \{tl\_cars=red $\wedge$ tl\_peds=green\})))), SUCCESS 
\end{framed}
\normalsize

\small
\begin{framed} 
\textbf{VO7}: LTL7.1 $\wedge$ LTL7.2
\end{framed}
\normalsize

The properties for \textbf{SAF1} - \textbf{SAF3} can be encoded as invariants.
Thus, they can be validated by an explicit-state model check, an LTL model check, or a symbolic model check.
In the following, we will validate them by VOs (\textbf{VO8} - \textbf{VO10}) applying respective explicit-state model checks (\textbf{MC1} - \textbf{MC3}).

Validation of \textbf{SAF1}:

\small
\begin{framed} 
\textbf{MC1}/TrafficLight/MC: $<$INV, tl\_cars = red or tl\_peds = red$>$
\end{framed}
\normalsize

\small
\begin{framed} 
\textbf{VO8}: MC1
\end{framed}
\normalsize

Validation of \textbf{SAF2}:

\small
\begin{framed} 
\textbf{MC2}/TrafficLight/MC: $<$INV, tl\_cars $\in$ \{red, redyellow, yellow, green\}$>$
\end{framed}
\normalsize

\small
\begin{framed} 
\textbf{VO9}: MC2
\end{framed}
\normalsize

Validation of \textbf{SAF3}:

\small
\begin{framed} 
\textbf{MC3}/TrafficLight/MC: $<$INV, tl\_peds $\in$ \{red, green\}$>$
\end{framed}
\normalsize

\small
\begin{framed} 
\textbf{VO10}: MC3
\end{framed}
\normalsize

In contrast, \textbf{LIV1} is a requirement describing a liveness property. Thus, it can be checked by \textbf{VO11} validating an LTL model check to expect a positive result as formalized in \textbf{LTL8}.

\small
\begin{framed} 
\textbf{LTL8}/TrafficLight/LTL: GF(\{tl\_cars = red $\wedge$ tl\_peds = red\}), SUCCESS
\end{framed}
\normalsize

\small
\begin{framed} 
\textbf{VO11}: LTL8
\end{framed}
\normalsize

For the validation of \textbf{SCENARIO1} and \textbf{SCENARIO2}, one needs (1) to replay them by executing the corresponding events, and (2)
to check the desired behavior afterwards.
To generate those scenarios, the modeler could animate the model, encode postconditions, and use the traces as regression tests afterwards.
This is realized in \textbf{TR1} and \textbf{TR2} respectively.
Afterwards, one could then define two respective VOs validating both VTs as shown in \textbf{VO12} and \textbf{VO13}.


\small
\begin{framed} 
\textbf{TR1}/TrafficLight/TR: [INITIALISATION $<$tl\_cars = red $\wedge$ tl\_peds = red$>$, cars\_ry $<$tl\_cars = redyellow $\wedge$ tl\_peds = red$>$, cars\_g $<$tl\_cars = green $\wedge$ tl\_peds = red\}, cars\_y $<$tl\_cars = yellow $\wedge$ tl\_peds = red$>$, cars\_r $<$tl\_cars = red $\wedge$ tl\_peds = red$>$]
\end{framed}
\normalsize

\small
\begin{framed} 
\textbf{VO12}: TR1
\end{framed}
\normalsize


\small
\begin{framed} 
\textbf{TR2}/TrafficLight/TR: [INITIALISATION $<$tl\_peds = red $\wedge$ tl\_cars = red$>$, peds\_g $<$tl\_peds = green $\wedge$ tl\_cars = red$>$, peds\_r $<$tl\_peds = red $\wedge$ tl\_cars = red$>$]
\end{framed}
\normalsize

\small
\begin{framed} 
\textbf{VO13}: TR2
\end{framed}
\normalsize

As mentioned before, a domain expert could be interested in sending commands only, without taking the traffic lights' colors into account.
The diagram portrayed in \textbf{PRC1} (\cref{projection-diagram}) could then be validated by projecting the state space on \texttt{queuedCmd} for inspection (formalized in \textbf{SPRJ1}) after applying explicit-state model checking to cover the complete state space (realized in \textbf{MC4}).
Both tasks are composed in \textbf{VO14} sequentially as \textbf{SPRJ1} depends on \textbf{MC4}.

\small
\begin{framed}
\textbf{MC4}/TrafficLight/MC: $<$FIN$>$
\end{framed}
\normalsize

\newpage

\small
\begin{framed} 
\textbf{SPRJ1}/TrafficLight\_Ref/SPRJ: queuedCmd, $S_{queuedCmd}$ = \{cmd\_none, cmd\_cars\_ry, cmd\_cars\_y, cmd\_cars\_g, cmd\_cars\_r, cmd\_peds\_r, cmd\_peds\_g\} $\wedge$ \\ 
$T_{queuedCmd}$ = \{(INITIALISATION, cmd\_none)\} $\cup$ \\
\{cmd\_none $\mapsto$ Send\_cmd(cars\_r) $\mapsto$ cmd\_cars\_r, \\ 
cmd\_none $\mapsto$ Send\_cmd(cars\_ry) $\mapsto$ cmd\_cars\_ry, \\
cmd\_none $\mapsto$ Send\_cmd(cars\_g) $\mapsto$ cmd\_cars\_g, \\ 
cmd\_none $\mapsto$ Send\_cmd(cars\_y) $\mapsto$ cmd\_cars\_y, \\ 
cmd\_none $\mapsto$ Send\_cmd(peds\_g) $\mapsto$ cmd\_peds\_g, \\ 
cmd\_none $\mapsto$ Send\_cmd(peds\_r) $\mapsto$ cmd\_peds\_r\} $\cup$ \\
\{cmd\_cars\_r $\mapsto$ Reject\_cmd $\mapsto$ cmd\_none, \\ 
cmd\_cars\_ry $\mapsto$ Reject\_cmd $\mapsto$ cmd\_none, \\ 
cmd\_cars\_g $\mapsto$ Reject\_cmd $\mapsto$ cmd\_none, \\ 
cmd\_cars\_y $\mapsto$ Reject\_cmd $\mapsto$ cmd\_none, \\ 
cmd\_peds\_g $\mapsto$ Reject\_cmd $\mapsto$ cmd\_none, \\ 
cmd\_peds\_r $\mapsto$ Reject\_cmd $\mapsto$ cmd\_none\} $\cup$ \\
\{cmd\_cars\_r $\mapsto$ cars\_r $\mapsto$ cmd\_none, cmd\_cars\_ry $\mapsto$ cars\_ry $\mapsto$ cmd\_none, \\ 
cmd\_cars\_g $\mapsto$ cars\_g $\mapsto$ cmd\_none, cmd\_cars\_y $\mapsto$ cars\_y $\mapsto$ cmd\_none, \\ 
cmd\_peds\_g $\mapsto$ peds\_g $\mapsto$ cmd\_none, cmd\_peds\_r $\mapsto$ peds\_r $\mapsto$ cmd\_none\}
\end{framed}
\normalsize

\small
\begin{framed} 
\textbf{VO14}: MC4;SPRJ1
\end{framed}
\normalsize


When validating \textbf{PROB-TIM1}, the modeler could apply hypothesis testing, or estimation of probability.
Here, we will demonstrate the validation of this requirement by applying hypothesis testing (see \textbf{HT1}).
The configuration to define a hypothesis depends on the tool.
In the context of \prob{}, in particular \simb{}, the VTs' parameters are represented as $(H, \alpha)$ with the hypothesis $H$, and the significance level $\alpha$.
Again, the hypothesis $H$ is represented as $(P_{start}, P_{end}, P_{prop}, T_{procedure}, Pr)$ containing the starting condition $P_{start}$, the ending condition $P_{end}$, the checked property $P_{prop}$, the test procedure $T_{procedure}$ and the probability $Pr$.
Here, the starting condition states that both traffic lights are red, the ending condition describes that 30 seconds have passed, the property to be checked is that the cars' traffic light eventually turns green, the kind of the hypothesis test is left-tailed, and the desired probability is 80 \%.
Again, the significance level is defined as 1\%.
Within the simulation configuration, we define 10000 runs to be simulated.
Validating \textbf{PROB-TIM2} is done similarly to \textbf{PROB-TIM1} with the main difference that the property to be checked states that the pedestrians' traffic light is green in the final state instead of the cars' traffic light (realized by \textbf{HT2}).
The respective VOs validating both VTs are shown in \textbf{VO15} and \textbf{VO16}.

\small
\begin{framed} 
\textbf{HT1}/TrafficLight, TrafficLight\_Sim/HT: ($<$PRED, tl\_cars = red $\wedge$ tl\_peds = red$>$, $<$TIME, 30000$>$, $<$EVENTUALLY, tl\_cars = green$>$, LEFT\_TAILED, 0.8), 0.01
\end{framed}
\normalsize

\small
\begin{framed} 
\textbf{VO15}: HT1
\end{framed}
\normalsize

\small
\begin{framed} 
\textbf{HT2}/TrafficLight, TrafficLight\_Sim/HT: ($<$PRED, tl\_cars = red $\wedge$ tl\_peds = red$>$, $<$TIME, 30000$>$, $<$EVENTUALLY, tl\_peds = green$>$, LEFT\_TAILED, 0.8), 0.01
\end{framed}
\normalsize

\small
\begin{framed} 
\textbf{VO16}: HT2
\end{framed}
\normalsize

In the following, we are going to demonstrate the validation of non-functional requirements.

As described before \textbf{COV1}, \textbf{COV2}, and \textbf{COV3} are coverage criteria for the validation of \textbf{SCENARIO1} and \textbf{SCENARIO2}.


In order to validate \textbf{COV1}, the modeler needs to inspect the variable coverage table after running \textbf{TR1} and \textbf{TR2} which validates \textbf{SCENARIO1} and \textbf{SCENARIO2} respectively. 
Here, it is necessary to check whether the values for \texttt{tl\_cars} and \texttt{tl\_peds} are equal to 4 and 2 respectively (realized in \textbf{VCT1}).

\small
\begin{framed}
\textbf{VCT1}/TrafficLight/VCT: $R_{vct}$(tl\_cars) = 4 $\wedge$ $R_{vct}$(tl\_peds) = 2
\end{framed}
\normalsize

\small
\begin{framed} 
\textbf{VO17}: (TR1 $\wedge$ TR2); VCT1
\end{framed}
\normalsize


Similar to the validation of \textbf{COV1}, the modeler must also run \textbf{TR1} and \textbf{TR2} validating \textbf{SCENARIO1} and \textbf{SCENARIO2} before validating \textbf{COV2}.
It is then necessary to inspect the operation coverage table manually, to check whether all events have been covered (realized by \textbf{OCT1}).

\small
\begin{framed}
\textbf{OCT1}/TrafficLight/OCT: \newline $\{$(cars\_ry, COVERED), (cars\_r, COVERED), (cars\_y, COVERED), \newline (cars\_r, COVERED), (peds\_r, COVERED), (peds\_g, COVERED)$\}$ = $R_{oct}$
\end{framed}
\normalsize

\small
\begin{framed} 
\textbf{VO18}: (TR1 $\wedge$ TR2); OCT1
\end{framed}
\normalsize


\textbf{COV3} describes the desired statistics for the number of states and transitions after running \textbf{TR1} and \textbf{TR2}, validating \textbf{SCENARIO1} and \textbf{SCENARIO2}.
As \textbf{SCENARIO1} and \textbf{SCENARIO2} should also cover the whole state space, the statistics are also expected to be equal to the statistics when applying explicit-state model checking. 
To check this coverage criterion, the modeler has to run \textbf{STAT1} afterwards.
Furthermore, \textbf{STAT1} has to be checked after running explicit-state model checking to cover the whole state space as shown in \textbf{MC4}.

\small
\begin{framed}
\textbf{STAT1}/TrafficLight/STAT: $R_{spstat}$("Number of States") = 6 $\wedge$ $R_{spstat}$("Number of Transitions") = 7
\end{framed}
\normalsize

\small
\begin{framed} 
\textbf{VO19}: ((TR1 $\wedge$ TR2); STAT1) $\wedge$ (MC4;STAT1)
\end{framed}
\normalsize


The requirements \textbf{STRUC1} and \textbf{STRUC2} desire \texttt{tl\_cars} and \texttt{tl\_peds} to be written by certain events.
This can be validated by the respective VOs \textbf{VO20} and \textbf{VO21}, running the respective tasks \textbf{RWM1} and \textbf{RWM2}.
Those VTs has to be checked by inspecting the read/write matrix manually.

\small
\begin{framed}
\textbf{RWM1}/TrafficLight/RWM: $R_{rwm}$[\{WRITE\}]$\sim$[\{tl\_cars\}] $=$ $\{$cars\_ry, cars\_g, cars\_y, cars\_r$\}$
\end{framed}
\normalsize

\small
\begin{framed} 
\textbf{VO20}: RWM1
\end{framed}
\normalsize


\small
\begin{framed}
\textbf{RWM2}/TrafficLight/RWM: $R_{rwm}$[\{WRITE\}]$\sim$[\{tl\_peds\}] $=$ $\{$peds\_g, peds\_r$\}$
\end{framed}
\normalsize

\small
\begin{framed} 
\textbf{VO21}: RWM2
\end{framed}
\normalsize


Again, ensuring that there are no vacuous parts in the invariant and guards of the Traffic Light model (\textbf{STRUC3}) can be checked by the VTs shown in \textbf{VAP1} and \textbf{VAP2}. The corresponding VO applying both tasks is realized in \textbf{VO22}.

\small
\begin{framed}
\textbf{VAP1}/TrafficLight/VAP: INV
\end{framed}
\normalsize

\small
\begin{framed}
\textbf{VAP2}/TrafficLight/VAP: GRD
\end{framed}
\normalsize

\small
\begin{framed} 
\textbf{VO22}: VAP1 $\wedge$ VAP2
\end{framed}
\normalsize


\textbf{STRUC4} describes how events enable each other.
This can be translated to task \textbf{ED1}.
The validation of \textbf{ED1} is done by \textbf{VO23}.

\small
\begin{framed}
\textbf{ED1}/TrafficLight/ED: \newline $\{$(cars\_ry, cars\_g), (cars\_g , cars\_y), (cars\_y , cars\_r), (cars\_r , cars\_ry), \newline  (cars\_r, peds\_g), (peds\_g , peds\_r), (peds\_r , peds\_g), (peds\_r, cars\_ry)$\} = $ $R_{ed}$.
\end{framed}
\normalsize

\small
\begin{framed} 
\textbf{VO23}: ED1
\end{framed}
\normalsize

For the validation of \textbf{STRUC5} and \textbf{STRUC6}, one could apply test case generation.
In order to validate \textbf{STRUC5}, test case generation covering all operations could be applied (realized by \textbf{OC1}).
Again, MCDC coverage test case generation is suitable to validate \textbf{STRUC6} (see \textbf{MCDC1}).
Afterwards, \textbf{STRUC5} and \textbf{STRUC6} are validated by \textbf{VO24} and \textbf{VO25} respectively.


\small
\begin{framed}
\textbf{OC1}/TrafficLight/OC:[cars\_r, cars\_ry, cars\_g, cars\_r, peds\_g, peds\_r]
\end{framed}
\normalsize

\small
\begin{framed} 
\textbf{VO24}: OC1
\end{framed}
\normalsize


\small
\begin{framed}
\textbf{MCDC1}/TrafficLight/MCDC:2
\end{framed}
\normalsize

\small
\begin{framed} 
\textbf{VO25}: MCDC1
\end{framed}
\normalsize

\paragraph{Other VOs to validate requirements}

Using the previous VOs, all requirements for the traffic light model have already been covered. 
In the following, we will now demonstrate VTs that have not been used yet. 
Some VTs will be demonstrated on existing requirements of the Traffic Light example.
In contrast, there will also be VTs that will be demonstrated on new requirements, or even other models.


Instead of validating \textbf{SAF1} by checking \textbf{MC1}, it would also be possible to apply symbolic model checking.
The corresponding VO, applying symbolic model checking (see \textbf{SMC1}) is shown in \textbf{VO26}.

\small
\begin{framed} 
\textbf{SMC1}/TrafficLight/SMC: $<$INV, tl\_cars = red or tl\_peds = red$>$
\end{framed}
\normalsize

\small
\begin{framed} 
\textbf{VO26}: SMC1
\end{framed}
\normalsize

Another possibility to validate \textbf{SAF1} could be done by proving multiple proof obligations.
Here, it would be necessary to generate a PO for the initialization, and for each event checking whether it preserves the invariant describing \textbf{SAF1}.
As result, this would lead to seven POs (denoted as \textbf{PO1} - \textbf{PO7}) being generated, one for each operation, to validate \textbf{SAF1}.
In order to fully validate \textbf{SAF1}, it is thus necessary to prove all POs as realized in \textbf{VO27}.
\textbf{PO1} shows the proof obligation (also used as validation task) for invariant preservation of the property describing \textbf{SAF1} from the event \texttt{cars\_ry}.
Note that proving POs might need some human interaction.

\small
\begin{framed} 
\textbf{PO1}/TrafficLight/PO: tl\_cars $\in$ colors, tl\_peds $\in$ \{red, green\}, tl\_peds = red or tl\_cars = red, tl\_cars = red, tl\_peds = red,
tl\_cars' = redyellow, tl\_peds' = tl\_peds $\models$ tl\_peds' = red or tl\_cars' = red
\end{framed}
\normalsize

\small
\begin{framed} 
\textbf{VO27}: PO1 $\wedge$ PO2 $\wedge$ $\ldots$ $\wedge$ PO7
\end{framed}
\normalsize


As an alternative to \textbf{STAT1}, one could also inspect the state space visualization (realized by \textbf{SVIS1}) after running \textbf{TR1} and \textbf{TR2}. The result could then be checked against the coverage after applying \textbf{MC4}. Both are realized in \textbf{VO28}.
As the state space can grow very fast, it is often better in practice to inspect the state space statistics after checking \textbf{MC4} as realized by \textbf{STAT1}.

\small
\begin{framed} 
\textbf{SVIS1}/TrafficLight/SVIS: $card(Z_{svis}) = 6 \wedge card(T_{svis}) = 7$
\end{framed}
\normalsize

\small
\begin{framed} 
\textbf{VO28}: ((TR1 $\wedge$ TR2); SVIS1) $\wedge$ (MC4; SVIS1)
\end{framed}
\normalsize


As an alternative to \textbf{LTL5.1} and \textbf{LTL7.1.}, it would also be possible to apply CTL model checking to expect a positive result.
The VOs applying the respective tasks \textbf{CTL1} and \textbf{CTL2} are shown in \textbf{VO29} and \textbf{VO30}.

\small
\begin{framed} 
\textbf{CTL1}/TrafficLight/CTL: EF\{tl\_cars $\neq$ red\}, SUCCESS
\end{framed}
\normalsize

\small
\begin{framed} 
\textbf{VO29}: CTL1
\end{framed}
\normalsize

\small
\begin{framed} 
\textbf{CTL2}/TrafficLight/CTL: EF\{tl\_peds $\neq$ red\}, SUCCESS
\end{framed}
\normalsize

\small
\begin{framed} 
\textbf{VO30}: CTL2
\end{framed}
\normalsize


Instead of applying hypothesis testing, it would also be possible to validate \textbf{PROB-TIM1} by estimating the probability.
The configuration for the check also depends on the tool.
Similar to hypothesis testing, the parameters contain the starting condition, the ending condition, the property to be checked, the kind of estimation checking, and the desired probability. The only difference is the $\delta$ value which is used instead of the $\alpha$ value. The corresponding VO is shown in \textbf{VO31}, validating the corresponding task \textbf{EOP1}.

\small
\begin{framed} 
\textbf{EOP1}/TrafficLight, TrafficLight\_Sim/EOP: 1000000, ($<$PRED, tl\_cars = red $\wedge$ tl\_peds = red$>$, $<$TIME, 30000$>$, $<$EVENTUALLY, tl\_cars = green$>$, LEFT\_TAILED, 0.8), 0.01
\end{framed}
\normalsize

\small
\begin{framed} 
\textbf{VO31}: EOP1
\end{framed}
\normalsize

Now, we will introduce a new requirement to demonstrate the VO for probabilistic/statistical model checking:

\small
\begin{framed} 
\textbf{PROB1:} Whenever both traffic lights are red, the pedestrians’ traffic light will turn green with a probability of 50\% next.
\end{framed}
\normalsize

In order to apply probabilistic/statistical model checking, the modeler has to encode a markov chain as well.
So, the demonstration of the corresponding VO is the only one that is not demonstrated using the B method and \prob{}.
An encoding of the Traffic Light model in PRISM is shown in \cref{lst:b-trafficlight-prism}.
This is also the context for the VT \textbf{PSMC1}.
Here, the probability to choose between the cars' cycle and the pedestrians' cycle is defined as 50\% for each.

\begin{lstlisting}[
caption=Traffic Light in PRISM,
label=lst:b-trafficlight-prism]
mdp

module TrafficLight_PRISM

    tl_cars : [0..3] init 0;
    tl_peds : [0..3] init 0;
   

    [] tl_cars=0 & tl_peds = 0 -> 0.5:(tl_cars'=1) + 0.5:(tl_peds'=2);
    [] tl_cars=1 -> (tl_cars' = 2);
    [] tl_cars=2 -> (tl_cars' = 3);
    [] tl_cars=3 -> (tl_cars' = 0);
    [] tl_peds=2 -> (tl_peds' = 0);

endmodule
\end{lstlisting}

Validating \textbf{PROB1} is then done by checking \textbf{VO32}, validating \textbf{PSMC1} with PCTL formula.

\small
\begin{framed} 
\textbf{PSMC1}/TrafficLight\_PRISM/PSMC: P$>$0.9999[$\neg$(true U ($\neg$((tl\_cars = 0 $\wedge$ tl\_peds = 0) $\implies$ (P$>$0.49 [ X (tl\_peds = 2) ] $\wedge$ P$<$0.51 [ X (tl\_peds = 2) ]))))], SUCCESS
\end{framed}
\normalsize

\small
\begin{framed} 
\textbf{VO32}: PSMC1
\end{framed}
\normalsize

\textbf{PROB1} could also be validated by inspecting the simulation statistics (realized in \textbf{SISTAT1}).
The corresponding VO is shown in \textbf{VO33}.

\small
\begin{framed} 
\textbf{SISTAT1}/TrafficLight,  TrafficLight\_Sim/SISTAT: 10000, $<$PRED, 1=1$>$, $<$STEPS, 100$>$, $R_{sistat}$(enabled $\mapsto$ peds\_g)/$R_{sistat}$(executed $\mapsto$ peds\_g) $\in$ [0.49, 0.51]
\end{framed}
\normalsize

\small
\begin{framed} 
\textbf{VO33}: SISTAT1
\end{framed}
\normalsize

As the Traffic Light model contains variables from the type \texttt{colors} only, it is not possible to inspect minimum and maximum values.
Let us consider a lift moving between the ground level and the 100th level.
Furthermore, assume that the level is modeled by a variable \texttt{level}.
Consider a scenario shown in \textbf{SCENARIO-LIFT}.

\small
\begin{framed} 
\textbf{SCENARIO-LIFT:}/In the beginning, the lift is located at the ground level. 
It then moves floor by floor until it reaches the third level.
\end{framed}
\normalsize

After validating \textbf{SCENARIO-LIFT}, i.e., after re-playing the scenario realized by a task \textbf{TR-LIFT}, it is expected that the lift has moved between the ground floor and the third level.
The corresponding requirement is shown in \textbf{COV-LIFT}.

\small
\begin{framed} 
\textbf{COV-LIFT:}/After validating \textbf{SCENARIO-LIFT}, it is expected that the lift has moved between the ground floor and the third level.
\end{framed}
\normalsize

\textbf{COV-LIFT} could then be validated by \textbf{VO-LIFT}, which inspects the minimum and maximum value as shown in \textbf{MMV-LIFT} after running \textbf{TR-LIFT}.

\small
\begin{framed} 
\textbf{MMV-LIFT}/Lift/MMV: min($R_{mmv}$(level)) = 0 $\wedge$ max($R_{mmv}$(level)) = 3
\end{framed}
\normalsize

\small
\begin{framed} 
\textbf{VO-LIFT}: TR-LIFT; MMV-LIFT
\end{framed}
\normalsize

Finally, we will also show a VO, applying model checking to search for a goal, which is used as an initial state to run trace replay.
This will be demonstrated in an automotive case study \cite{LeuschelMutz}.
Consider the scenario shown in \textbf{SCENARIO-AUTO}.

\small
\begin{framed}
\textbf{SCENARIO-AUTO:} \\
Assuming that the engine is turned on, and the blinker is in position \texttt{Downward7}. \\
After 500 ms, the lights on the left-hand side turn on with an intensity of 100. \\
When passing another 500ms, the lights on the left-hand side turn off. \\
Both events are repeated in the same order one more time. \\
While the lights on the left-hand side are blinking, those on the right-hand side are always turned off.
\end{framed}
\normalsize

First, the assumption of the scenario (engine turned on, and blinker in position \texttt{Downward7}) is validated by finding a state from which the other events of the scenario are executed. This is realized by the explicit-state model check \textbf{MC-AUTO}.
Outgoing from the state that should be found, the rest of the scenario is then validated via trace replay which is realized by \textbf{TR-AUTO}.

\small
\begin{framed}
\textbf{MC-AUTO}/PitmanController\_Time\_MC\_v4/MC: \\ $<$GOAL, engineOn $=$ TRUE $\wedge$ pitmanArmUpDown $=$ Downward7$>$
\end{framed}
\normalsize

\small
\begin{framed} 
\textbf{TR-AUTO}/PitmanController\_Time\_MC\_v4/TR:

[RTIME\_BlinkerOn(delta=500) $<$blinkLeft = 100, blinkRight=0$>$, \\ 
RTIME\_BlinkerOff(delta=500) $<$blinkLeft = 0, blinkRight=0$>$, \\
RTIME\_BlinkerOn(delta=500) $<$blinkLeft = 100, blinkRight=0$>$, \\ 
RTIME\_BlinkerOff(delta=500) $<$blinkLeft = 0, blinkRight=0$>$]
\end{framed}
\normalsize

\small
\begin{framed} 
\textbf{VO-AUTO}: MC-AUTO; TR-AUTO
\end{framed}
\normalsize

\begin{center}
\begin{figure}[ht]
    \centering
    \includegraphics[width=12cm]{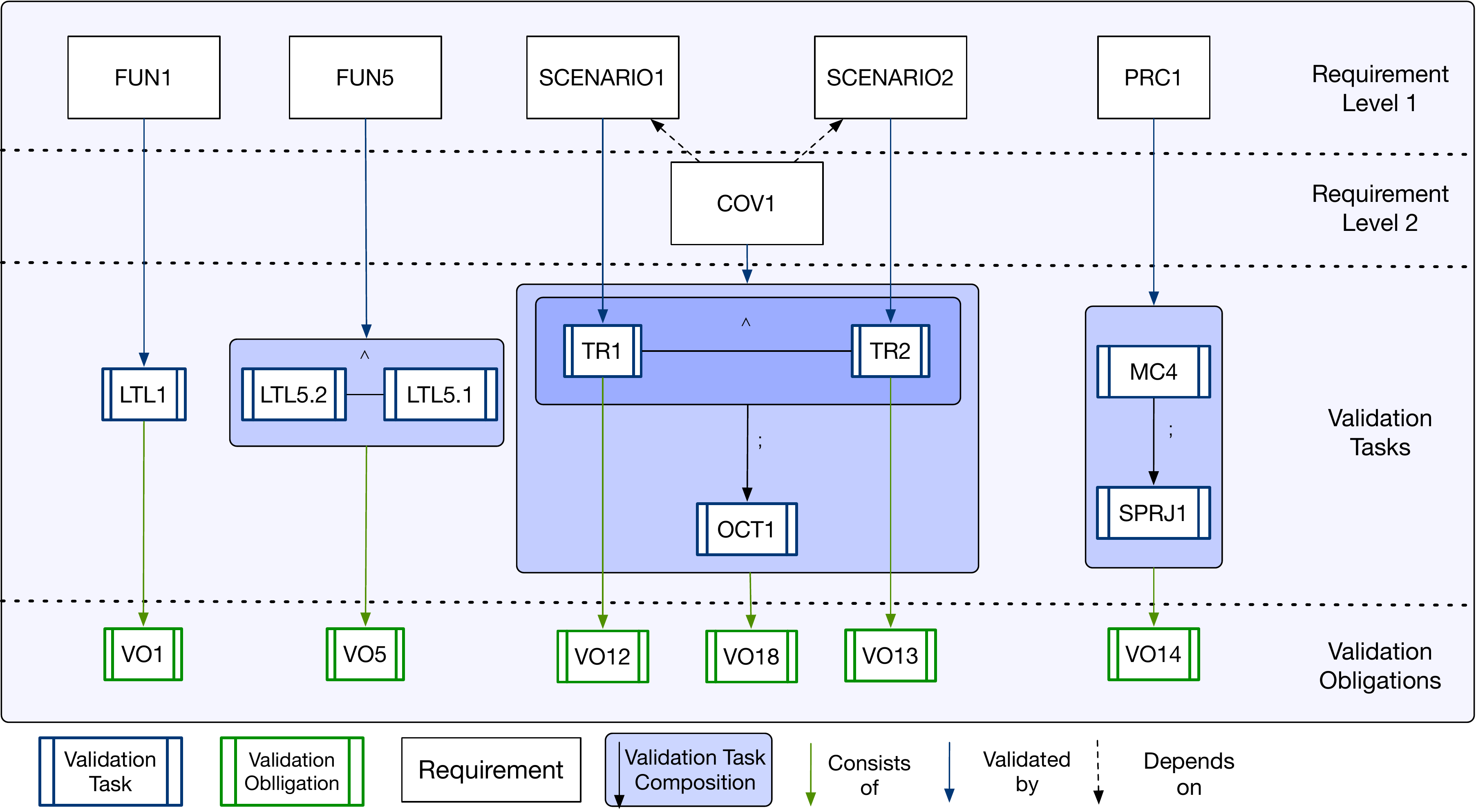}
    \caption{Taxonomy with Requirements FUN1, FUN5, SCENARIO1, SCENARIO2, COV1, PRC1, and corresponding VTs, and VOs validating them}
    \label{fig:taxonomy}
\end{figure}
\end{center}

\textbf{Traceability of Requirements} \cref{fig:taxonomy} shows a taxonomy of a subset of requirements (\textbf{FUN1}, \textbf{FUN5}, \textbf{SCENARIO1}, \textbf{SCENARIO2}, \textbf{COV1}, \textbf{PRC1}), as well as VOs, and VTs validating them.
Based on this taxonomy, we will demonstrate how possible error sources could be traced when a VO fails.
For example, assuming that $\mathbf{VO1}$ has failed, the possible error sources could be $\mathbf{LTL1}$, the requirement $\mathbf{FUN1}$, or the model.
Similarly, $\mathbf{VO5}$ is traced to $\mathbf{LTL5.1}$, $\mathbf{LTL5.2.}$, $\mathbf{FUN5}$, or the model.
Again, $\mathbf{VO14}$ is traced to $\mathbf{SPRJ1}$, $\mathbf{MC4}$, $\mathbf{PRC1}$, or the model.
While $\mathbf{VO12}$ and $\mathbf{VO13}$ are traced to $\mathbf{TR1}$ and $\mathbf{TR2}$, and thus also $\mathbf{SCENARIO1}$ and $\mathbf{SCENARIO2}$ respectively, tracing VOs to VTs and requirements is more complicated for $\mathbf{VO18}$.
Here, it is necessary to determine which parts of the VO has failed.
In the case that only $\mathbf{OCT_{1}}$ fails, it means that the set of scenarios might be incomplete.
In contrast, $\mathbf{TR3}$ and $\mathbf{TR4}$ are traced via $\mathbf{VO12}$ and $\mathbf{VO13}$ respectively.
This means that $\mathbf{VO18}$ could have failed because of the failure of $\mathbf{VO12}$ or $\mathbf{VO13}$.

In the case that an error is found, it is also important (1) to track contradicting requirements, and (2) to avoid introducing new bugs when trying to fix this error.
For example, fixing \textbf{SCENARIO1} in the model might cause other requirements to fail.
The modeler must then check, whether the \textbf{SCENARIO1} contradicts other requirements, or whether there are bugs introduced when the model evolves.

\appendix

\newpage

\section{Glossary}
\label{sec:glossary}

\paragraph{State}
The state of a (software) system is represented by the values of its variables (and constants).

\paragraph{Operation}
An operation is a term that is well-known from the formal B method. Analogous terms are, e.g., events or actions.
It consists of a guard, several effects, and optionally input and return parameters.
A guard is a predicate corresponding to an operation which is true when the operation is enabled.
When executing the operation, the effects are applied to the current state, modifying it to the succeeding state.

\paragraph{Transition}
A transition is labeled with an operation (and its parameters), and defined between two states $s_1$ and $s_2$ under the following condition:
The operation together with its parameters is enabled in $s_1$, and executing the operation with the parameters modifies $s_1$ resulting in $s_2$.

\paragraph{State Space}
A state space shows all possible executions of the system. It consists of a set of states and transitions between them.

\paragraph{Trace}
A trace is a list of transitions describing a path through the state space.

\paragraph{Scenario}
A scenario is a (high-level) sequence of events (written in natural language) that describes certain desired behavior patterns.

\paragraph{Differences between Scenario and Traces}
While researching literature it became apparent that the terms of \textit{trace} and \textit{scenario} have different meanings in the formal methods community.
Scenarios have also different meanings depending on the domain and context they are used in \cite{aiScenarios, scenariosInProcesses, uiScenarios, softwareScenarios}.
In the referenced paper it is somewhat agreed that a scenario describes a desired behavior. 
For software development, a scenario is then often expressed in a non-ambiguous DSL like Gherkin \cite{wynne2017cucumber}. 
In software engineering, there is the sentiment that traces are a realisation of a scenario, shown for example in \cite{scenarioAndTraces}. 
Depending on the underlying formalism a scenario can therefore have multiple traces that satisfy it.

\paragraph{Verification}
Verification checks whether a model meets its specification. So, here we ask the question: \textit{Are we building the software correctly?}.

\paragraph{Validation} 
Validation checks whether a model meets the stakeholder's requirements. So the main question is: \textit{Are we building the right software?}.

\paragraph{Validation Obligation}
A \emph{validation obligation} (VO) is composed of (multiple) validation tasks (VT) associated with a model to check its compliance with a requirement. Thus, validating a requirement succeeds if all associated VOs yield successful results.
Formally, a VO is annotated with an id, and consists of a \emph{validation expression} (written as: $\mathbf{VO_{id}}: \mathrm{VO_{expression}}$).

\paragraph{Validation Expression}
The \emph{validation expression} (VE) consists of operations on the associated validation tasks.
It is part of the validation obligation.

\paragraph{Validation Technique}
A validation technique is a technique to validate a requirement.
For example, one could validate a requirement describing a temporal property by LTL model checking.
In this case, LTL model checking is the validation technique.

\paragraph{Validation Task}
A validation task (VT) is identified with an identifier,
and consists of a validation technique that is applied with the given
validation parameters to the corresponding context.
Executing a VT possibly modifies the internal state of the validation tool, e.g., consisting of the currently explored state space, and the current trace.
The notation for a VT is as follows: $\mathbf{VT_{id}}$/$\mathrm{VT_{context}}$/$\mathrm{VT_{technique}}$: $\mathrm{VT_{parameters}}$

\newpage

\section{Traffic Light Refinement}

\begin{lstlisting}[
caption=Abstract Traffic Light,
label=lst:b-trafficlight-command-abstract]
MACHINE TrafficLight2
SETS colors = {red, redyellow, yellow, green};
     COMMANDS = {cmd_cars_ry, cmd_cars_y, cmd_cars_g, 
                 cmd_cars_r, cmd_peds_r, cmd_peds_g, cmd_none}
VARIABLES tl_cars, tl_peds
INVARIANT tl_cars : colors & tl_peds : {red, green} &
	  (tl_peds = red or tl_cars = red)
              
INITIALISATION  tl_cars := red || tl_peds := red
OPERATIONS
 Send_cmd(cmd) = SELECT cmd : COMMANDS THEN skip END;
 Reject_cmd = skip;

 cars_ry = SELECT tl_cars = red & tl_peds = red THEN tl_cars := redyellow END;
 cars_y = SELECT tl_cars = green THEN tl_cars := yellow END;
 cars_g = SELECT tl_cars = redyellow THEN tl_cars := green END; 
 cars_r = SELECT tl_cars = yellow THEN tl_cars := red END;
 peds_r = SELECT tl_peds = green THEN tl_peds := red END;
 peds_g = SELECT tl_cars = red & tl_peds = red THEN tl_peds := green END
END
\end{lstlisting}

\begin{lstlisting}[
caption=Traffic Light Refinement,
label=lst:b-trafficlight-command-concrete]
REFINEMENT TrafficLightCommand_Ref REFINES TrafficLight2
VARIABLES tl_cars, tl_peds, queuedCmd
INVARIANT queuedCmd : COMMANDS
INITIALISATION  tl_cars := red || tl_peds := red || queuedCmd := cmd_none
OPERATIONS
  Send_cmd(cmd) = 
    SELECT cmd : COMMANDS & cmd /= cmd_none & queuedCmd = cmd_none 
    THEN 
      queuedCmd := cmd 
    END;
  Reject_cmd = 
    SELECT queuedCmd /= cmd_none 
    THEN 
      queuedCmd := cmd_none 
    END;

   cars_ry = 
     SELECT 
       tl_cars = red & tl_peds = red & queuedCmd = cmd_cars_ry
     THEN 
       tl_cars := redyellow ||
       queuedCmd := cmd_none
     END;

   cars_y = 
     SELECT 
       tl_cars = green & queuedCmd = cmd_cars_y
     THEN 
      tl_cars := yellow ||
      queuedCmd := cmd_none
     END;

   cars_g = 
     SELECT 
       tl_cars = redyellow & queuedCmd = cmd_cars_g
     THEN 
       tl_cars := green ||
       queuedCmd := cmd_none
     END; 

   cars_r = 
     SELECT 
       tl_cars = yellow & queuedCmd = cmd_cars_r
     THEN 
       tl_cars := red ||
       queuedCmd := cmd_none
     END;

   peds_r = 
     SELECT 
       tl_peds = green & queuedCmd = cmd_peds_r
     THEN 
       tl_peds := red ||
       queuedCmd := cmd_none
     END;

   peds_g = 
      SELECT 
        tl_cars = red & tl_peds = red & queuedCmd = cmd_peds_g
      THEN 
        tl_peds := green ||
        queuedCmd := cmd_none
      END


END
\end{lstlisting}

\newpage

\section{Overview VT Examples}
\label{overview-vts}

Trace Replay VT:

\small
\begin{framed} 
\textbf{TR1}/TrafficLight/TR: [INITIALISATION $<$tl\_cars = red $\wedge$ tl\_peds = red$>$, cars\_ry $<$tl\_cars = redyellow $\wedge$ tl\_peds = red$>$, cars\_g $<$tl\_cars = green $\wedge$ tl\_peds = red\}, cars\_y $<$tl\_cars = yellow $\wedge$ tl\_peds = red$>$, cars\_r $<$tl\_cars = red $\wedge$ tl\_peds = red$>$]
\end{framed}
\normalsize

Operation Coverage Test Case Generation VT:

\small
\begin{framed}
\textbf{OC1}/TrafficLight/OC:[cars\_r, cars\_ry, cars\_g, cars\_r, peds\_g, peds\_r]
\end{framed}
\normalsize

MC/DC Coverage Test Case Generation VT:

\small
\begin{framed}
\textbf{MCDC1}/TrafficLight/MCDC:2
\end{framed}
\normalsize

Hypothesis Testing VT:

\small
\begin{framed} 
\textbf{HT1}/TrafficLight, TrafficLight\_Sim/HT: ($<$PRED, tl\_cars = red $\wedge$ tl\_peds = red$>$, $<$TIME, 30000$>$, $<$EVENTUALLY, tl\_cars = green$>$, LEFT\_TAILED, 0.8), 0.01
\end{framed}
\normalsize

Estimation of Probability VT:

\small
\begin{framed} 
\textbf{EOP1}/TrafficLight, TrafficLight\_Sim/EOP: ($<$PRED, tl\_cars = red $\wedge$ tl\_peds = red$>$, $<$TIME, 30000$>$, $<$EVENTUALLY, tl\_cars = green$>$, LEFT\_TAILED, 0.8), 0.01
\end{framed}
\normalsize

Simulation Statistics VT:

\small
\begin{framed} 
\textbf{SISTAT1}/TrafficLight,  TrafficLight\_Sim/SISTAT: $<$PRED, 1=1$>$, $<$STEPS, 100$>$, $R_{sistat}$(enabled $\mapsto$ cars\_ry)/$R_{sistat}$(executed $\mapsto$ cars\_ry) $\in$ [0.49, 0.51]
\end{framed}
\normalsize

Explicit-State Model Checking VT:

\small
\begin{framed} 
\textbf{MC1}/TrafficLight/MC: $<$INV, tl\_cars = red or tl\_peds = red$>$
\end{framed}
\normalsize

LTL Model Checking VT:

\small
\begin{framed} 
\textbf{LTL1}/TrafficLight/LTL: \{tl\_cars = red $\wedge$ tl\_peds = red\}, SUCCESS
\end{framed}
\normalsize

CTL Model Checking VT:

\small
\begin{framed} 
\textbf{CTL1}/TrafficLight/CTL: EF\{tl\_cars $\neq$ red\}, SUCCESS
\end{framed}
\normalsize

Probabilistic/Statistical Model Checking VT:

\small
\begin{framed} 
\textbf{PSMC1}/TrafficLight\_PRISM/PSMC: P$>$0.9999[$\neg$(true U ($\neg$((tl\_cars = 0 $\wedge$ tl\_peds = 0) $\implies$ (P$>$0.49 [ X (tl\_peds = 2) ] $\wedge$ P$<$0.51 [ X (tl\_peds = 2) ]))))], SUCCESS
\end{framed}
\normalsize

Symbolic Model Checking VT:

\small
\begin{framed} 
\textbf{SMC1}/TrafficLight/SMC: $<$INV, tl\_cars = red or tl\_peds = red$>$
\end{framed}
\normalsize

Proving VT:

\small
\begin{framed} 
\textbf{PO1}/TrafficLight/PO: tl\_cars $\in$ colors, tl\_peds $\in$ \{red, green\}, tl\_peds = red or tl\_cars = red, tl\_cars = red, tl\_peds = red,
tl\_cars' = redyellow, tl\_peds' = tl\_peds $\models$ tl\_peds' = red or tl\_cars' = red
\end{framed}
\normalsize

Variable Coverage Table VT:

\small
\begin{framed}
\textbf{VCT1}/TrafficLight/VCT: $R_{vct}$(tl\_cars) = 4 $\wedge$ $R_{vct}$(tl\_peds) = 2
\end{framed}
\normalsize

Min/Max Values VT:

\small
\begin{framed} 
\textbf{MMV-LIFT}/Lift/MMV: min($R_{mmv}$(level)) = 0 $\wedge$ max($R_{mmv}$(level)) = 3
\end{framed}
\normalsize

Operation Coverage Table VT:

\small
\begin{framed}
\textbf{OCT1}/TrafficLight/OCT: \newline $\{$(cars\_ry, COVERED), (cars\_r, COVERED), (cars\_y, COVERED), \newline (cars\_r, COVERED), (peds\_r, COVERED), (peds\_g, COVERED)$\}$ = $R_{oct}$
\end{framed}
\normalsize

Read/Write Matrix VT:

\small
\begin{framed}
\textbf{RWM1}/TrafficLight/RWM: $R_{rwm}$[\{WRITE\}]$\sim$[\{tl\_cars\}] $=$ $\{$cars\_ry, cars\_g, cars\_y, cars\_r$\}$
\end{framed}
\normalsize

Enabling Diagram VT:

\small
\begin{framed}
\textbf{ED1}/TrafficLight/ED: \newline $\{$(cars\_ry, cars\_g), (cars\_g , cars\_y), (cars\_y , cars\_r), (cars\_r , cars\_ry), \newline  (cars\_r, peds\_g), (peds\_g , peds\_r), (peds\_r , peds\_g), (peds\_r, cars\_ry)$\} = $ $R_{ed}$.
\end{framed}
\normalsize

Vacuous Parts VT:

\small
\begin{framed}
\textbf{VAP1}/TrafficLight/VAP: INV
\end{framed}
\normalsize

State Space Visualization VT:

\small
\begin{framed} 
\textbf{SVIS1}/TrafficLight/SVIS: $card(Z_{svis}) = 6 \wedge card(T_{svis}) = 7$
\end{framed}
\normalsize

State Space Projection VT:

\small
\begin{framed} 
\textbf{SPRJ1}/TrafficLight\_Ref/SPRJ: queuedCmd, $S_{queuedCmd}$ = \{$<$undefined$>$, cmd\_none, cmd\_cars\_ry, cmd\_cars\_y, cmd\_cars\_g, cmd\_cars\_r, cmd\_peds\_r, cmd\_peds\_g\} $\wedge$ \\ 
$T_{queuedCmd}$ = \{($<$undefined$>$, INITIALISATION, cmd\_none)\} $\cup$ \\
\{cmd\_none $\mapsto$ Send\_cmd(cars\_r) $\mapsto$ cmd\_cars\_r, \\ 
cmd\_none $\mapsto$ Send\_cmd(cars\_ry) $\mapsto$ cmd\_cars\_ry, \\
cmd\_none $\mapsto$ Send\_cmd(cars\_g) $\mapsto$ cmd\_cars\_g, \\ 
cmd\_none $\mapsto$ Send\_cmd(cars\_y) $\mapsto$ cmd\_cars\_y, \\ 
cmd\_none $\mapsto$ Send\_cmd(peds\_g) $\mapsto$ cmd\_peds\_g, \\ 
cmd\_none $\mapsto$ Send\_cmd(peds\_r) $\mapsto$ cmd\_peds\_r\} $\cup$ \\
\{cmd\_cars\_r $\mapsto$ Reject\_cmd $\mapsto$ cmd\_none, \\ 
cmd\_cars\_ry $\mapsto$ Reject\_cmd $\mapsto$ cmd\_none, \\ 
cmd\_cars\_g $\mapsto$ Reject\_cmd $\mapsto$ cmd\_none, \\ 
cmd\_cars\_y $\mapsto$ Reject\_cmd $\mapsto$ cmd\_none, \\ 
cmd\_peds\_g $\mapsto$ Reject\_cmd $\mapsto$ cmd\_none, \\ 
cmd\_peds\_r $\mapsto$ Reject\_cmd $\mapsto$ cmd\_none\} $\cup$ \\
\{cmd\_cars\_r $\mapsto$ cars\_r $\mapsto$ cmd\_none, cmd\_cars\_ry $\mapsto$ cars\_ry $\mapsto$ cmd\_none, \\ 
cmd\_cars\_g $\mapsto$ cars\_g $\mapsto$ cmd\_none, cmd\_cars\_y $\mapsto$ cars\_y $\mapsto$ cmd\_none, \\ 
cmd\_peds\_g $\mapsto$ peds\_g $\mapsto$ cmd\_none, cmd\_peds\_r $\mapsto$ peds\_r $\mapsto$ cmd\_none\} \\
\end{framed}
\normalsize

State Space Statistics VT:

\small
\begin{framed}
\textbf{STAT1}/TrafficLight/STAT: $R_{spstat}$("Number of States") = 6 $\wedge$ $R_{spstat}$("Number of Transitions") = 7
\end{framed}
\normalsize

\section{Published Papers}
\label{sec:publications}

In the following, we list the papers that are published in the context of D 1.1. of the IVOIRE project:

\begin{itemize}
\item Atif Mashkoor, Michael Leuschel, Alexander Egyed. Validation Obligations: A Novel Approach to Check Compliance between Requirements and their Formal Specification \cite{mashkoor2021validation}
\item Fabian Vu, Michael Leuschel, Atif Mashkoor. Validation of Formal Models by Timed Probabilistic Simulation \cite{vu2021validation}
\item Atif Mashkoor, Alexander Egyed. Evaluating the alignment of sequence diagrams with system behavior \cite{mashkoor2021evaluating}
\item Jens Bendisposto, David Gele\ss{}us, Yumiko Jansing, Michael Leuschel, Antonia P\"utz, Fabian Vu, Michelle Werth. ProB2-UI: A Java-Based User Interface for ProB \cite{prob2_ui} (extended in context of IVOIRE)
\end{itemize}

\section{Changes History}

\paragraph{Version 1.0.0:} Initial Version

\paragraph{Version 1.1.0:}

\begin{itemize}
\item Introduction updated
\item Classification of Requirements - now Section 2 instead of Section 4
\item "Overlap between Validation and Verification" renamed to "Verification and Validation"
\item "Verification and Validation" updated
\item "Comparison with Proof Obligations" (Section 2.1) removed - now discussed in Section 1
\item Validation Obligations Approach - now Section 4 instead of Section 2
\item Section 2.4. ("Refinement") and Section 2.5. ("Refactoring") - merged to "Refinement and Refactoring" and moved to Section 4.1.
\item Section 5 ("Validation Techniques and Validation Obligations") - now moved to Section 4.2. and renamed to "Validation Techniques and Tasks"
\item Separate \emph{validation obligation} and \emph{validation task}
\item Introduce definition for \emph{validation task} in "Validation Techniques and Tasks"
\item Introduce parallel ($\mid\mid$) and sequential (;) composition for validation task
\item Introduce Table 1 - showing overview of all validation tasks
\item Remove mathematical formulations of validation tasks - now described in Table 1
\item Remove Trace Refinement as validation task (originally Section 5.2.)
\item Minor changes in Section 4.2. (originally Section 5.)
\item Vacuous Guards/Invariants - not part of "Validation by Visualization, Statistics, and Metrics" anymore
\item Introduce Section 4.3. ("Definition of Validation Obligation")
\item Formal Definition of Validation Obligation updated
\item Minor changes in Formulation of "Definition of Validation Obligation"
\item Section 2.3 "Creating Validation Tasks" - now renamed to "Creating Validation Obligations" (Section 4.4.) 
\item Section 2.2. "Development" - now renamed to "VO-guided Workflow" and moved to Section 4.5.
\item Remove Figure 2 in "VO-guided Workflow" (originally Section 2.2.)
\item Section 6 "Demonstration of Validation Obligations" - now Section 5
\item Minor changes in formulations in "Demonstration of Validation Obligations"
\item Remove \textbf{PRC1} in "Demonstration of Validation Obligations"
\item Separation between \emph{validation task} and \emph{validation obligation} in "Demonstration of Validation Obligations" following the new definitions in Section 4.2. "Validation Techniques and Tasks" and Section 4.3. "Definition of Validation Obligation"
\item Use parallel and sequential composition of \emph{validation task} in VOs in "Demonstration of Validation Obligations"
\item Remove \textbf{TRF1} validating \textbf{PRC1} in "Demonstration of Validation Obligations"
\item Add and Explain Figure 3: "Taxonomy with Requirements FUN1, FUN5, SCENARIO1, SCENARIO2, COV1, PRC1, and corresponding VTs, and VOs validating them"
\item Rename \textbf{MMV1} to \textbf{MMV-LIFT}
\item Minor changes in formulations in Glossary
\item Update \emph{validation task} in Glossary
\item Add \emph{validation obligation}, \emph{validation predicate}, and \emph{validation function} in Glossary
\item Rename Appendix C "Overview VO Examples" to "Overview VT Examples"
\item Rename VO to VT in "Overview VT Examples"
\end{itemize}

\paragraph{Version 1.2.0:}

\begin{itemize}
\item Section 1 ("Introduction") updated
\item Minor changes in Section 2.2 ("User Requirements and System Requirements")
\item Section 3 ("Verification and Validation") updated, Figure 1 ("Role of Verification and Validation") added
\item Minor changes in introduction of Section 4 ("Validation Obligations Approach")
\item Minor changes in Section 4.1 ("Refinement and Refactoring")
\item Split Section 4.2 ("Validation Techniques and Tasks") in Section 4.2 ("Definition of Validation Tasks") and Section 4.3 ("Classification of Validation Techniques")
\item Table 1 moved to the end of Section 4.3 ("Classification of Validation Techniques")
\item Table 1 changed (Number of Simulations for \textbf{HT} and \textbf{EOP} removed, Depth for \textbf{OC} and \textbf{MCDC} removed, style changed)
\item Operators applied on VTs are now part of the VO, in particular, the $VO_{expression}$, and not part of the VT anymore
\item Minor changes in introduction of Section 4.3 ("Classification of Validation Techniques")
\item Minor changes in "Validation by Visualization, Statistics, and Metrics" and "Code Generation for Validation" of Section 4.3 ("Classification of Validation Techniques")
\item Minor change in textual definition of a VO (Section 4.4: "Definition of Validation Obligations")
\item Replace $VO_{predicate}$ by $VO_{expression}$ in formal definition of a VO (Section 4.4: "Definition of Validation Obligations")
\item Remove validation function (Section 4.4: "Definition of Validation Obligations")
\item Adopt explaination of operations in $VO_{expression}$ to formal definition
\item Remove parallel operator ($\mid\mid$) as it has the same semantics as the and operator ($\wedge$)
\item Minor changes in Section 4.5 ("Creating Validation Obligations")
\item Minor change/improvement in \textbf{FUN8} in Section 5 ("Demonstration of Validation Obligations")
\item Split Section 5 ("Demonstration of Validation Obligations") in two paragraphs: \textbf{Requirements} and \textbf{Validation by VOs}
\item Remove \texttt{validate} predicate in \textbf{Validation by VOs}, Section 5 ("Demonstration of Validation Obligations")
\item Replace $\&$ by $\wedge$ in all VOs in Section 5 ("Demonstration of Validation Obligations") and Appendix
\item Minor changes in text of \textbf{Validation by VOs}, Section 5 ("Demonstration of Validation Obligations")
\item Remove $<$undefined$>$ from \textbf{SPRJ1}
\item Minor change in text for validation of \textbf{PROB-TIM1}, \textbf{COV1}, \textbf{COV3}, \textbf{STRUC1}, and \textbf{STRUC2}
\item Change \textbf{STRUC5} and \textbf{STRUC6}
\item Remove depth in \textbf{OC1} and \textbf{MCDC1}
\item Minor change in \textbf{SCENARIO-AUTO}
\item Split discussion with Figure 4 ("Taxonomy with Requirements FUN1, FUN5, SCENARIO1, SCENARIO2, COV1, PRC1, and corresponding VTs, and VOs validating them") to separate paragraph in Section 5: "Traceability of Requirements"
\item Updated Figure 4 ("Taxonomy with Requirements FUN1, FUN5, SCENARIO1, SCENARIO2, COV1, PRC1, and corresponding VTs, and VOs validating them") 
\item More texts added in Section 5: "Traceability of Requirements"
\item Remove \emph{Validation Function} and \emph{Validation Predicate} in Glossary
\item Add \emph{Validation Expression} in Glossary
\item Replace \emph{Validation Predicate} by \emph{Validation Expression} in \emph{Validation Obligation} in Glossary
\item Minor changes in formulations in Glossary: \emph{validation task}, \emph{scenario}
\item Remove Composed VTs in Appendix D ("Overview VT Examples")
\item Appendix A ("Changes history") moved to Appendix E
\end{itemize}

\newpage

\listoffigures
\listoftables

\addcontentsline{toc}{section}{List of Listings}

\lstlistoflistings

\printbibliography

@inproceedings{lin2015towards,
	title        = {Towards verifying VDM using SPIN},
	author       = {Lin, Hsin-Hung and Omori, Yoichi and Kusakabe, Shigeru and Araki, Keijiro},
	year         = 2015,
	booktitle    = {International Workshop on Formal Techniques for Safety-Critical Systems},
	pages        = {241--256},
	organization = {Springer}
}

@inproceedings{yu1999model,
	title        = {Model checking TLA+ specifications},
	author       = {Yu, Yuan and Manolios, Panagiotis and Lamport, Leslie},
	year         = 1999,
	booktitle    = {Advanced Research Working Conference on Correct Hardware Design and Verification Methods},
	pages        = {54--66},
	organization = {Springer}
}

@inproceedings{derrick2008z2sal,
	title        = {Z2SAL-building a model checker for Z},
	author       = {Derrick, John and North, Siobh{\'a}n and Simons, Anthony JH},
	year         = 2008,
	booktitle    = {International Conference on Abstract State Machines, B and Z},
	pages        = {280--293},
	organization = {Springer}
}

@inproceedings{gibson2014fdr3,
	title        = {FDR3--a modern refinement checker for CSP},
	author       = {Gibson-Robinson, Thomas and Armstrong, Philip and Boulgakov, Alexandre and Roscoe, Andrew W},
	year         = 2014,
	booktitle    = {International Conference on Tools and Algorithms for the Construction and Analysis of Systems},
	pages        = {187--201},
	organization = {Springer}
}

@article{plagge2010seven,
	title        = {Seven at one stroke: LTL model checking for high-level specifications in B, Z, CSP, and more},
	author       = {Plagge, Daniel and Leuschel, Michael},
	year         = 2010,
	journal      = {International journal on software tools for technology transfer},
	publisher    = {Springer},
	volume       = 12,
	number       = 1,
	pages        = {9--21}
}

@inproceedings{vakili2012temporal,
	title        = {Temporal logic model checking in Alloy},
	author       = {Vakili, Amirhossein and Day, Nancy A},
	year         = 2012,
	booktitle    = {International Conference on Abstract State Machines, Alloy, B, VDM, and Z},
	pages        = {150--163},
	organization = {Springer}
}

@inproceedings{leuschel2003prob,
	title        = {ProB: A model checker for B},
	author       = {Leuschel, Michael and Butler, Michael},
	year         = 2003,
	booktitle    = {International symposium of formal methods europe},
	pages        = {855--874},
	organization = {Springer}
}

@phdthesis{krings2017towards,
	title        = {Towards infinite-state symbolic model checking for B and Event-B},
	author       = {Krings, Sebastian},
	year         = 2017,
	school       = {Universit{\"a}ts-und Landesbibliothek der Heinrich-Heine-Universit{\"a}t D{\"u}sseldorf}
}

@book{ulbrich2011proving,
	title        = {{}On proving alloy specifications using KeY}
}

@inproceedings{gargantini2007using,
	title        = {Using model checking to generate fault detecting tests},
	author       = {Gargantini, Angelo},
	year         = 2007,
	booktitle    = {International Conference on Tests and Proofs},
	pages        = {189--206},
	organization = {Springer}
}

@phdthesis{freitas2005model,
	title        = {Model checking circus},
	author       = {Freitas, Leonardo},
	year         = 2005,
	school       = {University of York}
}

@inproceedings{plagge2007validating,
	title        = {Validating Z specifications using the ProB animator and model checker},
	author       = {Plagge, Daniel and Leuschel, Michael},
	year         = 2007,
	booktitle    = {International Conference on Integrated Formal Methods},
	pages        = {480--500},
	organization = {Springer}
}

@book{abrial2005b,
	title        = {The B-book: assigning programs to meanings},
	author       = {Abrial, Jean-Raymond},
	year         = 2005,
	publisher    = {Cambridge university press}
}

@article{abrial2007rodin,
	title        = {{Rodin: An Open Toolset for Modelling and Reasoning in Event-B}},
	author       = {Abrial, Jean-Raymond and Butler, Michael and Hallerstede, Stefan and Hoang, Thai Son and Mehta, Farhad and Voisin, Laurent},
	year         = 2010,
	month        = nov,
	journal      = {Int. J. Softw. Tools Technol. Transf.},
	publisher    = {Springer-Verlag},
	address      = {Berlin, Heidelberg},
	volume       = 12,
	number       = 6,
	pages        = {447–466},
	issn         = {1433-2779},
	issue_date   = {November 2010},
	numpages     = 20
}

@article{borger2003asm,
	title        = {The ASM refinement method},
	author       = {B{\"o}rger, Egon},
	year         = 2003,
	journal      = {Formal aspects of computing},
	publisher    = {Springer},
	volume       = 15,
	number       = 2,
	pages        = {237--257}
}

@inproceedings{mentre2012discharging,
	title        = {Discharging proof obligations from Atelier B using multiple automated provers},
	author       = {Mentr{\'e}, David and March{\'e}, Claude and Filli{\^a}tre, Jean-Christophe and Asuka, Masashi},
	year         = 2012,
	booktitle    = {International Conference on Abstract State Machines, Alloy, B, VDM, and Z},
	pages        = {238--251},
	organization = {Springer}
}

@inproceedings{agerholm1996translating,
	title        = {Translating specifications in VDM-SL to PVS},
	author       = {Agerholm, Sten},
	year         = 1996,
	booktitle    = {International Conference on Theorem Proving in Higher Order Logics},
	pages        = {1--16},
	organization = {Springer}
}

@inproceedings{maharaj1997verification,
	title        = {On the verification of VDM specification and refinement with PVS},
	author       = {Maharaj, Savi and Bicarregui, Juan},
	year         = 1997,
	booktitle    = {Proceedings 12th IEEE International Conference Automated Software Engineering},
	pages        = {280--289},
	organization = {IEEE}
}

@inproceedings{stringer1997using,
	title        = {Using PVS to prove a Z refinement: A case study},
	author       = {Stringer-Calvert, David WJ and Stepney, Susan and Wand, Ian},
	year         = 1997,
	booktitle    = {International Symposium of Formal Methods Europe},
	pages        = {573--588},
	organization = {Springer}
}

@inproceedings{bonfanti2017asm2c++,
	title        = {Asm2C++: a tool for code generation from abstract state machines to Arduino},
	author       = {Bonfanti, Silvia and Carissoni, Marco and Gargantini, Angelo and Mashkoor, Atif},
	year         = 2017,
	booktitle    = {NASA Formal Methods Symposium},
	pages        = {295--301},
	organization = {Springer}
}

@article{rivera2017code,
	title        = {Code generation for Event-B},
	author       = {Rivera, Victor and Catano, N{\'e}stor and Wahls, Tim and Rueda, Camilo},
	year         = 2017,
	journal      = {International Journal on Software Tools for Technology Transfer},
	publisher    = {Springer},
	volume       = 19,
	number       = 1,
	pages        = {31--52}
}

@inproceedings{vu2019multi,
	title        = {A multi-target code generator for high-level B},
	author       = {Vu, Fabian and Hansen, Dominik and K{\"o}rner, Philipp and Leuschel, Michael},
	year         = 2019,
	booktitle    = {International Conference on Integrated Formal Methods},
	pages        = {456--473},
	organization = {Springer}
}

@article{hasanagic2019code,
	title        = {Code generation for distributed embedded systems with VDM-RT},
	author       = {Hasanagi{\'c}, Miran and Fabbri, Tommaso and Larsen, Peter Gorm and Bandur, Victor and Tran-J{\o}rgensen, Peter and Ouy, Julien},
	year         = 2019,
	journal      = {Design Automation for Embedded Systems},
	publisher    = {Springer},
	volume       = 23,
	number       = 3,
	pages        = {153--177}
}

@online{alloyAnalyzer,
	title        = {The Alloy Analyzer},
	url          = {https://alloytools.org/},
	urldate      = {2021-06-16}
}

@article{snook2021domain,
	title        = {Domain-specific scenarios for refinement-based methods},
	author       = {Snook, Colin and Hoang, Thai Son and Dghaym, Dana and Fathabadi, Asieh Salehi and Butler, Michael},
	year         = 2021,
	journal      = {Journal of Systems Architecture},
	publisher    = {Elsevier},
	volume       = 112,
	pages        = 101833
}

@inproceedings{droschl1999design,
	title        = {Design and application of a test case generator for VDM-SL},
	author       = {Droschl, Georg},
	year         = 1999,
	booktitle    = {Workshop Materials: VDM in Practice}
}

@inproceedings{hansen2012translating,
	title        = {Translating TLA+ to B for validation with ProB},
	author       = {Hansen, Dominik and Leuschel, Michael},
	year         = 2012,
	booktitle    = {International Conference on Integrated Formal Methods},
	pages        = {24--38},
	organization = {Springer}
}

@inproceedings{arcaini2016smt,
	title        = {SMT-based automatic proof of ASM model refinement},
	author       = {Arcaini, Paolo and Gargantini, Angelo and Riccobene, Elvinia},
	year         = 2016,
	booktitle    = {International Conference on Software Engineering and Formal Methods},
	pages        = {253--269},
	organization = {Springer}
}

@inproceedings{helke1997automating,
	title        = {Automating test case generation from Z specifications with Isabelle},
	author       = {Helke, Steffen and Neustupny, Thomas and Santen, Thomas},
	year         = 1997,
	booktitle    = {International conference of Z users},
	pages        = {52--71},
	organization = {Springer}
}

@inproceedings{vu2021validation,
	title        = {Validation of Formal Models by Timed Probabilistic Simulation},
	author       = {Vu, Fabian and Leuschel, Michael and Mashkoor, Atif},
	year         = 2021,
	booktitle    = {International Conference on Rigorous State-Based Methods},
	pages        = {81--96},
	organization = {Springer}
}

@inproceedings{testGenerationCSP,
	title        = {Guided Test Generation from CSP Models},
	author       = {Nogueira, Sidney and Sampaio, Augusto and Mota, Alexandre},
	year         = 2008,
	booktitle    = {Theoretical Aspects of Computing - ICTAC 2008},
	publisher    = {Springer Berlin Heidelberg},
	address      = {Berlin, Heidelberg},
	editor       = {Fitzgerald, John S. and Haxthausen, Anne E. and Yenigun, Husnu}
}

@inproceedings{CoSimulationVDM,
	title        = {Collaborative Modelling and Co-simulation in the Development of Dependable Embedded Systems},
	author       = {Fitzgerald, John and Larsen, Peter Gorm and Pierce, Ken and Verhoef, Marcel and Wolff, Sune},
	year         = 2010,
	booktitle    = {Integrated Formal Methods},
	publisher    = {Springer Berlin Heidelberg},
	address      = {Berlin, Heidelberg},
	editor       = {M{\'e}ry, Dominique and Merz, Stephan}
}

@inproceedings{visB,
	title        = {VisB: A Lightweight Tool to Visualize Formal Models with SVG Graphics},
	author       = {Werth, Michelle and Leuschel, Michael},
	year         = 2020,
	booktitle    = {Rigorous State-Based Methods},
	publisher    = {Springer International Publishing},
	address      = {Cham},
	editor       = {Raschke, Alexander and M{\'e}ry, Dominique and Houdek, Frank}
}

@manual{clearsy2016atelierb,
	title        = {{Atelier {B}, User and Reference Manuals}},
	author       = {ClearSy},
	year         = 2016,
	address      = {Aix-en-Provence, France},
	note         = {Available at \url{http://www.atelierb.eu/}}
}

@article{snook2006uml,
	title        = {UML-B: Formal modeling and design aided by UML},
	author       = {Snook, Colin and Butler, Michael},
	year         = 2006,
	journal      = {ACM Transactions on Software Engineering and Methodology (TOSEM)},
	publisher    = {ACM New York, NY, USA},
	volume       = 15,
	number       = 1,
	pages        = {92--122}
}

@book{modelchecking,
	title        = {{Principles of Model Checking}},
	author       = {Christel Baier and Joost-Pieter Katoen},
	year         = 2008,
	publisher    = {MIT Press}
}

@misc{role_validation,
	title        = {{The Role of Validation in Refinement-Based Formal Software Development}},
	author       = {Jacquot, Jean-Pierre and Mashkoor, Atif},
	year         = 2018,
	journal      = {{Models: Concepts, Theory, Logic, Reasoning, and Semantics - Essays Dedicated to Klaus-Dieter Schewe on the Occasion of His 60th Birthday}},
	volume       = 34,
	pages        = {202--219}
}

@inproceedings{DBLP:conf/icfem/LadenbergerL15,
	title        = {Mastering the Visualization of Larger State Spaces with Projection Diagrams},
	author       = {Lukas Ladenberger and Michael Leuschel},
	year         = 2015,
	booktitle    = {Proceedings {ICFEM} 2015},
	series       = {LNCS 9407},
	pages        = {153--169},
	doi          = {10.1007/978-3-319-25423-4\_10},
	url          = {https://doi.org/10.1007/978-3-319-25423-4\%5F10},
	bibsource    = {dblp computer science bibliography, https://dblp.org}
}

@inproceedings{smc1,
	title        = {{Statistical Model Checking: An Overview}},
	author       = {Legay, Axel and Delahaye, Beno{\^i}t and Bensalem, Saddek},
	year         = 2010,
	booktitle    = {Runtime Verification},
	series       = {LNCS},
	volume       = 6418
}

@inbook{smc2,
	title        = {Statistical Model Checking},
	author       = {Legay, Axel and Lukina, Anna and Traonouez, Louis Marie and Yang, Junxing and Smolka, Scott A. and Grosu, Radu},
	year         = 2019,
	booktitle    = {Computing and Software Science: State of the Art and Perspectives},
	publisher    = {Springer International Publishing},
	address      = {Cham},
	series       = {LNCS},
	volume       = 10000,
	pages        = {478--504},
	doi          = {10.1007/978-3-319-91908-9\_23},
	isbn         = {978-3-319-91908-9},
	url          = {https://doi.org/10.1007/978-3-319-91908-9\%5F23}
}

@article{fisher_1925,
	title        = {Theory of Statistical Estimation},
	author       = {Fisher, Ronald A.},
	year         = 1925,
	journal      = {Mathematical Proceedings of the Cambridge Philosophical Society},
	publisher    = {Cambridge University Press},
	volume       = 22,
	number       = 5,
	pages        = {700–725},
	doi          = {10.1017/S0305004100009580}
}

@inproceedings{hypothesistest,
	title        = {{Kendall's Advanced Theory of Statistics}},
	author       = {Kendall, Maurice G. and Stuart, Alan and Ord, J. Keith},
	year         = 1987,
	booktitle    = {Oxford University Press},
	isbn         = {0195205618}
}

@inproceedings{montecarlo,
	title        = {{Monte Carlo Simulation}},
	author       = {Mooney, Christopher Z.},
	year         = 1997,
	booktitle    = {Sage publications},
	volume       = 116
}

@article{derrick2011z2sal,
	title        = {Z2SAL: a translation-based model checker for Z},
	author       = {Derrick, John and North, Siobh{\'a}n and Simons, Anthony JH},
	year         = 2011,
	journal      = {Formal Aspects of Computing},
	publisher    = {Springer},
	volume       = 23,
	number       = 1,
	pages        = {43--71}
}

@inproceedings{konnov2018bmcmt,
	title        = {BmcMT: Bounded Model Checking of TLA+ Specifications with SMT},
	author       = {Konnov, Igor and Kukovec, Jure and Tran, Thanh},
	year         = 2018,
	booktitle    = {TLA+ Community Meeting 2018}
}

@article{konnov2019tla+,
	title        = {TLA+ model checking made symbolic},
	author       = {Konnov, Igor and Kukovec, Jure and Tran, Thanh-Hai},
	year         = 2019,
	journal      = {Proceedings of the ACM on Programming Languages},
	publisher    = {ACM New York, NY, USA},
	volume       = 3,
	number       = {OOPSLA},
	pages        = {1--30}
}

@inproceedings{symbolicZ,
	title        = {Model Checking Z Specifications Using SAL},
	author       = {Smith, Graeme and Wildman, Luke},
	year         = 2005,
	booktitle    = {ZB 2005: Formal Specification and Development in Z and B},
	publisher    = {Springer Berlin Heidelberg},
	address      = {Berlin, Heidelberg},
	pages        = {85--103},
	isbn         = {978-3-540-32007-4}
}

@book{event-b-probabilistic-1,
	title        = {Reliability Assessment in Event-B Development},
	author       = {Anton Tarasyuk, Elena Troubitsyna, Linas Laibinis},
	year         = 2009,
	booktitle    = {NODES 09: NOrdic workshop and doctoral symposium on DEpendability and Security, Link{\"o}ping, Sweden, April 27, 2009},
	publisher    = {Link{\"o}ping electronic Press}
}

@article{event-b-probabilistic-2,
	title        = {From Formal Specification in Event-B to Probabilistic Reliability Assessment},
	author       = {Tarasyuk, Anton and Troubitsyna, Elena and Laibinis, Linas},
	year         = 2010,
	month        = {07},
	journal      = {Dependability, International Conference on},
	pages        = {24--31},
	doi          = {10.1109/DEPEND.2010.12},
	isbn         = {978-0-7695-4090-0}
}

@inproceedings{prob2,
	title        = {{ProB 2.0 Tutorial}},
	author       = {{Bendisposto, Jens and Clark, Joy and Dobrikov, Ivaylo and K\"{o}rner, Philipp and Krings, Sebastian and Ladenberger, Lukas and Leuschel, Michael and Plagge Daniel}},
	year         = 2013,
	booktitle    = {{Proceedings of the 4th Rodin User and Developer Workshop}},
	publisher    = {TUCS Lecture Notes}
}

@inproceedings{ProBCSP,
	title        = {Combining CSP and B for Specification and Property Verification},
	author       = {Butler, Michael and Leuschel, Michael},
	year         = 2005,
	booktitle    = {FM 2005: Formal Methods},
	publisher    = {Springer Berlin Heidelberg},
	address      = {Berlin, Heidelberg},
	pages        = {221--236},
	isbn         = {978-3-540-31714-2}
}

@misc{mashkoor2021validation,
	title        = {Validation Obligations: A Novel Approach to Check Compliance between Requirements and their Formal Specification},
	author       = {Atif Mashkoor and Michael Leuschel and Alexander Egyed},
	year         = 2021,
	eprint       = {2102.06037},
	archiveprefix = {arXiv},
	primaryclass = {cs.SE}
}

@inproceedings{barrocas2012jcircus,
	title        = {JCircus 2.0: an Extension of an Automatic Translator from Circus to Java.},
	author       = {Barrocas, Samuel Lincoln Magalh{\~a}es and Oliveira, Marcel},
	year         = 2012,
	booktitle    = {CPA},
	pages        = {15--36}
}

@inproceedings{oda2015vdm,
	title        = {VDM animation for a wider range of stakeholders},
	author       = {Oda, Tomohiro and Yamomoto, Y and Nakakoji, Kumiyo and Araki, Keijiro and Larsen, Peter Gorm},
	year         = 2015,
	booktitle    = {Proceedings of the 13th Overture Workshop},
	pages        = {18--32},
	organization = {Citeseer}
}

@inproceedings{sullivan2017automated,
	title        = {Automated test generation and mutation testing for Alloy},
	author       = {Sullivan, Allison and Wang, Kaiyuan and Zaeem, Razieh Nokhbeh and Khurshid, Sarfraz},
	year         = 2017,
	booktitle    = {2017 IEEE International Conference on Software Testing, Verification and Validation (ICST)},
	pages        = {264--275},
	organization = {IEEE}
}

@inproceedings{gargantini2003using,
	title        = {Using Spin to generate tests from ASM specifications},
	author       = {Gargantini, Angelo and Riccobene, Elvinia and Rinzivillo, Salvatore},
	year         = 2003,
	booktitle    = {International Workshop on Abstract State Machines},
	pages        = {263--277},
	organization = {Springer}
}

@inproceedings{testcase_1,
	title        = {Model-Based Robustness Testing in Event-B Using Mutation},
	author       = {Savary, Aymerick and Frappier, Marc and Leuschel, Michael and Lanet, Jean-Louis},
	year         = 2015,
	booktitle    = {Software Engineering and Formal Methods},
	publisher    = {Springer International Publishing},
	address      = {Cham},
	pages        = {132--147},
	isbn         = {978-3-319-22969-0},
	editor       = {Calinescu, Radu and Rumpe, Bernhard}
}

@inproceedings{testcase_2,
	title        = {Applying Model Checking to Generate Model-Based Integration Tests from Choreography Models},
	author       = {Wieczorek, Sebastian and Kozyura, Vitaly and Roth, Andreas and Leuschel, Michael and Bendisposto, Jens and Plagge, Daniel and Schieferdecker, Ina},
	year         = 2009,
	booktitle    = {Testing of Software and Communication Systems},
	publisher    = {Springer Berlin Heidelberg},
	address      = {Berlin, Heidelberg},
	pages        = {179--194},
	isbn         = {978-3-642-05031-2},
	editor       = {N{\'u}{\~{n}}ez, Manuel and Baker, Paul and Merayo, Mercedes G.}
}

@inproceedings{asmetaa,
	title        = {AsmetaA: Animator for Abstract State Machines},
	author       = {Bonfanti, Silvia and Gargantini, Angelo and Mashkoor, Atif},
	year         = 2018,
	booktitle    = {Abstract State Machines, Alloy, B, TLA, VDM, and Z},
	publisher    = {Springer International Publishing},
	address      = {Cham},
	pages        = {369--373},
	isbn         = {978-3-319-91271-4},
	editor       = {Butler, Michael and Raschke, Alexander and Hoang, Thai Son and Reichl, Klaus}
}

@inproceedings{asmetav,
	title        = {A Scenario-Based Validation Language for ASMs},
	author       = {Carioni, Alessandro and Gargantini, Angelo and Riccobene, Elvinia and Scandurra, Patrizia},
	year         = 2008,
	booktitle    = {Abstract State Machines, B and Z},
	publisher    = {Springer Berlin Heidelberg},
	address      = {Berlin, Heidelberg},
	pages        = {71--84},
	isbn         = {978-3-540-87603-8},
	editor       = {B{\"o}rger, Egon and Butler, Michael and Bowen, Jonathan P. and Boca, Paul}
}

@inproceedings{cspSolving,
	title        = {Azucar: A SAT-Based CSP Solver Using Compact Order Encoding},
	author       = {Tanjo, Tomoya and Tamura, Naoyuki and Banbara, Mutsunori},
	year         = 2012,
	booktitle    = {Theory and Applications of Satisfiability Testing -- SAT 2012},
	publisher    = {Springer Berlin Heidelberg},
	address      = {Berlin, Heidelberg},
	pages        = {456--462},
	isbn         = {978-3-642-31612-8}
}

@inproceedings{asmetamc,
	title        = {AsmetaSMV: A Way to Link High-Level ASM Models to Low-Level NuSMV Specifications},
	author       = {Arcaini, Paolo and Gargantini, Angelo and Riccobene, Elvinia},
	year         = 2010,
	booktitle    = {Abstract State Machines, Alloy, B and Z},
	publisher    = {Springer Berlin Heidelberg},
	address      = {Berlin, Heidelberg},
	pages        = {61--74},
	isbn         = {978-3-642-11811-1},
	editor       = {Frappier, Marc and Gl{\"a}sser, Uwe and Khurshid, Sarfraz and Laleau, R{\'e}gine and Reeves, Steve}
}

@article{Gargantini2001ASMBasedTC,
	title        = {ASM-Based Testing: Coverage Criteria and Automatic Test Sequence},
	author       = {A. Gargantini and E. Riccobene},
	year         = 2001,
	journal      = {J. Univers. Comput. Sci.},
	volume       = 7,
	pages        = {1050--1067}
}

@inproceedings{prob2_ui,
	title        = {{ProB2-UI: A Java-based User Interface for ProB}},
	author       = {Bendisposto, Jens and Gele\ss{}us, David and Jansing, Yumiko and Leuschel, Michael and P\"utz, Antonia and Vu, Fabian and Werth, Michelle},
	year         = 2021,
	booktitle    = {Proceedings FMICS},
	organization = {Springer}
}

@misc{chaudhuri2008tla,
	title        = {A TLA+ Proof System},
	author       = {Kaustuv C. Chaudhuri and Damien Doligez and Leslie Lamport and Stephan Merz},
	year         = 2008,
	eprint       = {0811.1914},
	archiveprefix = {arXiv},
	primaryclass = {cs.LO}
}

@article{DBLP:journals/simpra/ThuleLGML19,
	title        = {Maestro: The {INTO-CPS} co-simulation framework},
	author       = {Casper Thule and Kenneth Lausdahl and Cl{\'{a}}udio Gomes and Gerd Meisl and Peter Gorm Larsen},
	year         = 2019,
	journal      = {Simul. Model. Pract. Theory},
	volume       = 92,
	pages        = {45--61},
	doi          = {10.1016/j.simpat.2018.12.005},
	url          = {https://doi.org/10.1016/j.simpat.2018.12.005},
	bibsource    = {dblp computer science bibliography, https://dblp.org}
}

@inproceedings{csp_spin,
	title        = {How to make FDR spin: LTL model checking of CSP by refinement},
	author       = {Leuschel, Michael and Massart, Thierry and Currie, Andrew},
	year         = 2001,
	month        = {01},
	journal      = {LNCS},
	volume       = 2021,
	pages        = {99--118}
}

@article{ideStdVal,
	title        = {Validation},
	year         = 1991,
	journal      = {IEEE Std 610},
	pages        = {1--217},
	doi          = {10.1109/IEEESTD.1991.106963}
}

@article{ideStdVerif,
	title        = {Verification},
	year         = 1991,
	journal      = {IEEE Std 610},
	pages        = {1--217},
	doi          = {10.1109/IEEESTD.1991.106963}
}

@article{enabling,
	title        = {Enabling analysis for Event-B},
	author       = {Dobrikov, Ivaylo and Leuschel, Michael},
	year         = 2017,
	month        = {08},
	journal      = {Science of Computer Programming},
	doi          = {10.1016/j.scico.2017.08.004}
}

@inproceedings{aiScenarios,
	title        = {{ViZDoom: A Doom-based AI research platform for visual reinforcement learning}},
	author       = {Kempka, Micha\l{} and Wydmuch, Marek and Runc, Grzegorz and Toczek, Jakub and Ja\'{s}kowski, Wojciech},
	year         = 2016,
	booktitle    = {2016 IEEE Conference on Computational Intelligence and Games (CIG)},
	pages        = {1--8},
	doi          = {10.1109/CIG.2016.7860433}
}

@article{scenariosInProcesses,
	title        = {{End-to-end Automatic Business Process Validation}},
	author       = {Ana C.R. Paiva and Nuno H. Flores and Jo\~{a}o P. Faria and Jos\'{e} M.G. Marques},
	year         = 2018,
	journal      = {Procedia Computer Science},
	volume       = 130,
	pages        = {999--1004},
	doi          = {https://doi.org/10.1016/j.procs.2018.04.104},
	issn         = {1877-0509},
	url          = {https://www.sciencedirect.com/science/article/pii/S1877050918304666},
	note         = {The 9th International Conference on Ambient Systems, Networks and Technologies (ANT 2018) / The 8th International Conference on Sustainable Energy Information Technology (SEIT-2018) / Affiliated Workshops}
}

@article{uiScenarios,
	title        = {{Task-centered user interface design}},
	author       = {Lewis, Clayton and Rieman, John},
	year         = 1993,
	journal      = {A practical introduction}
}

@inproceedings{softwareScenarios,
	title        = {{A scenario-based approach to validating and testing software systems using statecharts}},
	author       = {Ryser, Johannes and Glinz, Martin},
	year         = 1999,
	booktitle    = {Proc. 12th International Conference on Software and Systems Engineering and their Applications},
	organization = {Citeseer}
}

@inproceedings{scenarioAndTraces,
	title        = {{Detecting Implied Scenarios from Execution Traces}},
	author       = {Mendonca, Nabor C. and Uchitel, Sebastian and Kramer, Jeff and Cantal de Sousa, Felipe},
	year         = 2007,
	booktitle    = {14th Working Conference on Reverse Engineering (WCRE 2007)},
	pages        = {50--59},
	doi          = {10.1109/WCRE.2007.19}
}

@book{wynne2017cucumber,
	title        = {{The cucumber book: behaviour-driven development for testers and developers}},
	author       = {Wynne, Matt and Hellesoy, Aslak and Tooke, Steve},
	year         = 2017,
	publisher    = {Pragmatic Bookshelf}
}

@article{asmetas,
	title        = {A Metamodel-based Language and a Simulation Engine for Abstract State Machines},
	author       = {Gargantini, Angelo and Riccobene, Elvinia and Scandurra, Patrizia},
	year         = 2008,
	month        = {01},
	journal      = {Journal of Universal Computer Science},
	volume       = 14,
	pages        = {1949--1983}
}

@article{brunel2019simulation,
	title        = {Simulation under arbitrary temporal logic constraints},
	author       = {Brunel, Julien and Chemouil, David and Cunha, Alcino and Macedo, Nuno},
	year         = 2019,
	journal      = {arXiv preprint arXiv:1912.10634}
}

@article{Kuppe_2019,
	title        = {{The TLA+ Toolbox}},
	author       = {Kuppe, Markus Alexander and Lamport, Leslie and Ricketts, Daniel},
	year         = 2019,
	month        = {Dec},
	journal      = {Electronic Proceedings in Theoretical Computer Science},
	publisher    = {Open Publishing Association},
	volume       = 310,
	pages        = {50–62},
	doi          = {10.4204/eptcs.310.6},
	issn         = {2075-2180},
	url          = {http://dx.doi.org/10.4204/EPTCS.310.6}
}

@article{z_replay,
	title        = {Integration of sequential scenarios},
	author       = {Desharnais, J. and Frappier, M. and Khedri, R. and Mili, A.},
	year         = 1998,
	journal      = {IEEE Transactions on Software Engineering},
	volume       = 24,
	number       = 9,
	pages        = {695--708},
	doi          = {10.1109/32.713325}
}

@inproceedings{carter2007mise,
	title        = {Mise en scene: converting scenarios to CSP traces in support of requirements-based programming},
	author       = {Carter, J and Gardner, William B},
	year         = 2007,
	booktitle    = {31st IEEE Software Engineering Workshop (SEW 2007)},
	pages        = {41--52},
	organization = {IEEE}
}

@inbook{electrum,
	title        = {Validating the Hybrid ERTMS/ETCS Level 3 Concept with Electrum},
	author       = {Cunha, Alcino and Macedo, Nuno},
	year         = 2018,
	month        = {01},
	pages        = {307--321},
	doi          = {10.1007/978-3-319-91271-4\_21},
	isbn         = {978-3-319-91270-7},
	howpublished = {\url{https://github.com/haslab/Electrum/releases/tag/v1.0}}
}

@manual{prob_test_case_generation,
	title        = {The ProB Animator and Model Checker Wiki}
}

@incollection{isobe2008proof,
	title        = {Proof principles of CSP--CSP-Prover in practice},
	author       = {Isobe, Yoshinao and Roggenbach, Markus},
	year         = 2008,
	booktitle    = {Dynamics in Logistics},
	publisher    = {Springer},
	pages        = {425--442}
}

@inproceedings{sun2008bounded,
	title        = {Bounded model checking of compositional processes},
	author       = {Sun, Jun and Liu, Yang and Dong, Jin Song and Sun, Jing},
	year         = 2008,
	booktitle    = {2008 2nd IFIP/IEEE International Symposium on Theoretical Aspects of Software Engineering},
	pages        = {23--30},
	organization = {IEEE}
}

@article{taibi2009stepwise,
	title        = {Stepwise Refinement Validation of Design Patterns Formalized in TLA+ using the TLC Model Checker.},
	author       = {Taibi, Toufik and Herranz-Nieva, Angel and Moreno-Navarro, Juan Jos{\'e}},
	year         = 2009,
	journal      = {J. Object Technol.},
	volume       = 8,
	number       = 2,
	pages        = {137--161}
}

@article{zproof,
	title        = {HOL-Z 2.0: A proof environment for Z-specifications},
	author       = {Brucker, Achim and Rittinger, Frank and Wolff, Burkhart},
	year         = 2003,
	month        = {01},
	journal      = {J. UCS},
	volume       = 9,
	pages        = {152--172}
}

@inproceedings{Freitas2005ProvingTW,
	title        = {Proving Theorems with Z / Eves},
	author       = {L. Freitas},
	year         = 2005
}

@inproceedings{brunel2018electrum,
	title        = {The electrum analyzer: model checking relational first-order temporal specifications},
	author       = {Brunel, Julien and Chemouil, David and Cunha, Alcino and Macedo, Nuno},
	year         = 2018,
	booktitle    = {Proceedings of the 33rd ACM/IEEE International Conference on Automated Software Engineering},
	pages        = {884--887}
}

@inproceedings{kiv,
	title        = {Formal System Development with KIV},
	author       = {Balser, Michael and Reif, Wolfgang and Schellhorn, Gerhard and Stenzel, Kurt and Thums, Andreas},
	year         = 2000,
	month        = {03},
	volume       = 1783,
	pages        = {363--366},
	doi          = {10.1007/3-540-46428-X\_25},
	isbn         = {978-3-540-67261-6}
}

@inproceedings{LeuschelMutz,
	title        = {{Modelling and Validating an Automotive System in Classical B and Event-B}},
	author       = {Leuschel, Michael and Mutz, Mareike and Werth, Michelle},
	year         = 2020,
	booktitle    = {Proceedings ABZ},
	series       = {LNCS},
	pages        = {335--350}
}

@article{ieee729,
	title        = {IEEE Standard Glossary of Software Engineering Terminology},
	year         = 1983,
	journal      = {ANSI/ IEEE Std 729-1983},
	pages        = {1--40},
	doi          = {10.1109/IEEESTD.1983.7435207}
}

@book{Sommerville10,
	title        = {Software Engineering},
	author       = {Sommerville, Ian},
	year         = 2010,
	publisher    = {Addison-Wesley},
	address      = {Harlow, England},
	edition      = 9,
	chapter      = 4
}

@article{mashkoor2021evaluating,
author = {Mashkoor, Atif and Egyed, Alexander},
year = {2021},
month = {01},
pages = {502-506},
title = {Evaluating the alignment of sequence diagrams with system behavior},
volume = {180},
journal = {Procedia Computer Science},
doi = {10.1016/j.procs.2021.01.267}
}

@techreport{Rushby93:FAA,
    AUTHOR = {John Rushby},
    TITLE = {Formal Methods and the Certification of Critical Systems},
    INSTITUTION = {Computer Science Laboratory, {SRI} International},
    YEAR = {1993},
    NUMBER = {{SRI-CSL-93-7}},
    ADDRESS = {Menlo Park, {CA}},
    URL = {http://www.csl.sri.com/papers/csl-93-7/}
}

@InProceedings{spear,
author="Fifarek, Aaron W.
and Wagner, Lucas G.
and Hoffman, Jonathan A.
and Rodes, Benjamin D.
and Aiello, M. Anthony
and Davis, Jennifer A.",
title={{SpeAR} v2.0: Formalized {P}ast {LTL} Specification and Analysis of Requirements},
booktitle="Proceedings NFM",
series       = {LNCS 10227},
year="2017",
pages="420--426",
isbn="978-3-319-57288-8"
}

@inproceedings{kaos,
  title={The {KAOS} Project: Knowledge Acquisition in Automated Specification of Software},
  booktitle={Proceedings AAAI Spring Symposium Series},
  author={Axel van Lamsweerde and Anne Dardenne and Bruno Delcourt and Françoise Dubisy},
  year={1991},
  pages = {59--62}
}

@manual{doors,
	title        = {Engineering Requirements Management {DOORS} Overview},
	author       = {IBM},
	year         = 2021,
	note         = {Available at \url{https://www.ibm.com/docs/en/ermd/9.7.2?topic=engineering-requirements-management-doors-overview}}
}

@book{problem_frames,
    author = {Jackson, Michael},
    year = {2001},
    title = {Problem Frames: Analyzing and Structuring Software Development Problems},
    publisher = {Addison-Wesley}
}

@phdthesis{Jastram2012ThePA,
  title={The {P}ro{R} Approach: Traceability of Requirements and System Descriptions},
  author={Michael Jastram},
  school={Heinrich-Heine-Universit{\"a}t D{\"u}sseldorf},
  year={2012}
}

@inproceedings{DBLP:conf/asm/HansenL0KKNNS18,
  author    = {Dominik Hansen and
               Michael Leuschel and
               David Schneider and
               Sebastian Krings and
               Philipp K{\"{o}}rner and
               Thomas Naulin and
               Nader Nayeri and
               Frank Skowron},
  title     = {Using a Formal {B} Model at Runtime in a Demonstration of the {ETCS}
               Hybrid Level 3 Concept with Real Trains},
  booktitle = {Proceedings {ABZ}},
  pages     = {292--306},
  year      = {2018},
  url       = {https://doi.org/10.1007/978-3-319-91271-4\_20},
  doi       = {10.1007/978-3-319-91271-4\_20},
  bibsource = {dblp computer science bibliography, https://dblp.org}
}

@inproceedings{DBLP:conf/rssrail/ComptierLMPM19,
  author    = {Mathieu Comptier and
               Michael Leuschel and
               Luis{-}Fernando Mejia and
               Julien Molinero Perez and
               Mareike Mutz},
  title     = {Property-Based Modelling and Validation of a {CBTC} Zone Controller
               in {Event-B}},
  booktitle = {Proceedings RSSRail},
  pages     = {202--212},
  year      = {2019},
  url       = {https://doi.org/10.1007/978-3-030-18744-6\_13},
  doi       = {10.1007/978-3-030-18744-6\_13},
  bibsource = {dblp computer science bibliography, https://dblp.org}
}

@article{landinggear,
author = {Ladenberger, Lukas and Hansen, Dominik and Wiegard, Harald and Bendisposto, Jens and Leuschel, Michael},
title = {Validation of the {ABZ} Landing Gear System Using {P}ro{B}},
year = {2017},
publisher = {Springer-Verlag},
address = {Berlin, Heidelberg},
volume = {19},
number = {2},
issn = {1433-2779},
url = {https://doi.org/10.1007/s10009-015-0395-9},
doi = {10.1007/s10009-015-0395-9},
journal = {International  Journal  on  Software  Tools  for  Technology  Transfer},
month = apr,
pages = {187–203},
numpages = {17}
}

@article{mashkoor17a,
  author    = {Atif Mashkoor and
               Jean{-}Pierre Jacquot},
  title     = {Validation of formal specifications through transformation and animation},
  journal   = {Requir. Eng.},
  volume    = {22},
  number    = {4},
  pages     = {433--451},
  year      = {2017},
  url       = {https://doi.org/10.1007/s00766-016-0246-6},
  doi       = {10.1007/s00766-016-0246-6}
}

@article{mashkoor17b,
  author    = {Atif Mashkoor and
               Faqing Yang and
               Jean{-}Pierre Jacquot},
  title     = {Refinement-based Validation of Event-B Specifications},
  journal   = {Softw. Syst. Model.},
  volume    = {16},
  number    = {3},
  pages     = {789--808},
  year      = {2017},
  url       = {https://doi.org/10.1007/s10270-016-0514-4},
  doi       = {10.1007/s10270-016-0514-4}
}

@article{bonfanti20a,
  author    = {Silvia Bonfanti and
               Angelo Gargantini and
               Atif Mashkoor},
  title     = {Design and validation of a {C++} code generator from Abstract State
               Machines specifications},
  journal   = {J. Softw. Evol. Process.},
  volume    = {32},
  number    = {2},
  year      = {2020},
  url       = {https://doi.org/10.1002/smr.2205},
  doi       = {10.1002/smr.2205}
}

@article{mashkoor18a,
  author    = {Atif Mashkoor and
               Felix Kossak and
               Alexander Egyed},
  title     = {Evaluating the suitability of state-based formal methods for industrial
               deployment},
  journal   = {Softw. Pract. Exp.},
  volume    = {48},
  number    = {12},
  pages     = {2350--2379},
  year      = {2018},
  url       = {https://doi.org/10.1002/spe.2634},
  doi       = {10.1002/spe.2634}
}

\end{document}